\documentclass[12pt,preprint]{aastex}
\usepackage{natbib}

\newcommand{\lea}{{\>\rlap{\raise2pt\hbox{$<$}}\lower3pt\hbox{$\sim$} \>}}
\newcommand{\gea}{{\>\rlap{\raise2pt\hbox{$>$}}\lower3pt\hbox{$\sim$} \>}}

\shorttitle{CLUSTER DISRUPTION IN THE MAGELLANIC CLOUDS}
\shortauthors{Chandar et al.}

\usepackage{natbib}
\begin{document}

\title{NEW TESTS FOR DISRUPTION MECHANISMS\\
OF STAR CLUSTERS:\\
THE LARGE AND SMALL MAGELLANIC CLOUDS}

\author{Rupali Chandar,\altaffilmark{1}
       S.\ Michael Fall,\altaffilmark{2} \&
       Bradley C.\ Whitmore\altaffilmark{2}
}

\altaffiltext{1}{Department of Physics \& Astronomy,
         University of Toledo, Toledo, OH 43606;
         Rupali.Chandar@ utoledo.edu}
\altaffiltext{2}{Space Telescope Science Institute,
         3700 San Martin Drive, Baltimore, MD 21218;
         fall@stsci.edu, whitmore@stsci.edu}

\begin{abstract}

We compare the observed bivariate distribution of masses ($M$) and ages ($\tau$) of star clusters in the Large Magellanic Cloud (LMC) with the predicted distributions $g(M,\tau)$ from three idealized models for the disruption of star clusters: (1)~sudden mass-dependent disruption; 
(2)~gradual mass-dependent disruption; and (3)~gradual mass-independent
disruption. The model with mass-{\em in}dependent disruption provides a good, first-order description of these cluster populations, with $g(M,\tau) \propto M^{\beta}\tau^{\gamma}$, $\beta=-1.8\pm0.2$ and $\gamma=-0.8\pm0.2$, at least for clusters with ages $\tau \lea10^9$~yr and masses $M\gea10^3\,M_{\odot}$ (more specifically, $\tau \lea 10^7(M/10^2\,M_{\odot})^{1.3}$~yr). This model predicts that the clusters should have a power-law luminosity function, $dN/dL \propto L^{-1.8}$, in agreement with observations. The first two models, on the other hand, fare poorly when describing the observations, refuting previous claims that mass-dependent disruption of star clusters is observed in the LMC over the studied $M$--$\tau$ domain.  Clusters in the SMC can be described by the same $g(M,\tau)$ distribution as for the LMC, but with smaller samples and hence larger uncertainties. The successful $g(M,\tau)$ model for clusters in the Magellanic Clouds is virtually the same as the one for clusters in the merging Antennae galaxies, but extends the domain of validity to lower masses and to older ages. This indicates that the dominant disruption processes are similar in these very different galaxies over at least $\tau \lea 10^8$~yr and possibly $\tau \lea 10^9$~yr. The mass functions for young clusters in the LMC are power-laws, while that for ancient globular clusters is peaked. We show that the observed shapes of these mass functions are consistent with expectations from the simple evaporation model presented by McLaughlin \& Fall.

\end{abstract}


\section{INTRODUCTION}

The primary window into the formation and disruption of star clusters comes from their mass and age distributions. Some physical processes, such as the evaporation of stars due to internal relaxation, are known to disrupt low-mass clusters earlier than high-mass clusters. Any such mass-dependent disruption process will imprint features (such as a bend) in the mass and age distributions, $\psi(M)\propto dN/dM$ and $\chi(\tau)\propto dN/d\tau$, 
and will show up as correlations in the joint distribution of cluster masses and ages $g(M,\tau)$. We have recently derived new formulae for $g(M,\tau)$ for three idealized models, two in which the rate of disruption depends on the mass of a cluster, and one in which it does not:
(1)~sudden mass-dependent disruption, 
(2)~gradual mass-dependent disruption, and 
(3)~gradual mass-{\em in}dependent disruption.
A comparison of the predictions from these models with the observations of star clusters in the merging Antennae galaxies shows that gradual mass-independent disruption (Model~3) provides a good, first-order description of these clusters, with $g(M,\tau) \propto \psi(M) \chi(\tau) \propto M^{-2}\,\tau^{-1}$ over the studied range $\tau \lea 10^7 (M/10^4\,M_{\odot})^{1.3}$~yr (Fall, Chandar, \& Whitmore 2009; hereafter FCW09).
Results plotted in Lada \& Lada (2003) for young clusters in the solar neighborhood suggest $\psi(M)$ and $\chi(\tau)$ distributions that are similar to those for the Antennae, although they are noisier and pertain to clusters with lower masses than in the Antennae.

The main motivation for the present work is to test whether or not any of the disruption models mentioned above also describe young ($\tau \lea 10^9$~yr) star clusters in more normal galaxies than the Antennae. To accomplish this, we apply the same methodology that we used in the Antennae, to the star cluster systems in the Large and Small Magellanic Clouds (LMC and SMC). These neighboring galaxies provide an important test case because their clusters are relatively easy to identify and study, and they are representative of low-mass, late-type galaxies commonly found
throughout the nearby universe. The Magellanic Clouds are also in a more quiescent phase than the Antennae, and have had a nearly constant rate of star formation (to within a factor of two) for the last few Gyr (Harris \& Zaritsky 2004; 2009).

We use catalogs of star clusters  in the Magellanic Clouds provided by Hunter et~al.\ (2003), which are the most comprehensive and photometrically uniform samples currently available, to derive the mass and age for each cluster.\footnote{Hunter et~al.\ selected clusters to be highly concentrated aggregates of stars with a density higher than that of the surrounding stellar field, regardless of whether the clusters might be classified as open, populous or globular, and without assessing if the clusters are gravitationally bound, since this is virtually impossible to determine for clusters younger than $\sim$10 internal crossing times.} The mass and age distributions of the clusters  are then compared with $g(M,\tau)$ predictions from the three disruption models mentioned above. We follow the methodology developed in FCW09, which plots relatively narrow projections of $g(M,\tau)$ averaged over $\tau$ and $M$ and which lends itself to easy graphical interpretation. This simple but {\em direct} method simplifies the search for possible features that change over time, such as a bend in the mass function, as required when mass-dependent disruption affects a cluster system. Previously, an {\em indirect} approach developed by Boutloukos \& Lamers (2003) was used to infer values for the disruption timescale for star clusters in the LMC and SMC. Their method relies on mass and age distributions averaged over wide ranges of $M$ and $\tau$, and makes the {\em a priori} assumption that clusters are disrupted in a mass-dependent fashion. Here, we revisit the disruption of star clusters in the Magellanic Clouds using our more direct approach, and assess the validity of the previous results.

As mentioned above, there have been several earlier studies aimed at determining the mass and age distributions of star clusters in the LMC and SMC, and interpreting these distributions in terms of the formation and
disruption of the clusters. The recent literature on this subject is somewhat confusing and full of contradictory claims. We have examined these papers carefully, and find that despite the apparent conflicts, there is reasonably good agreement on the shape of the cluster mass and age distributions among many previous works, and also with the results presented here.
This is particularly true for the LMC. We summarize the results from these previous works and how they relate to ours in an accompanying Appendix.

The remainder of this paper is organized as follows: Section~2 provides a brief summary of the three idealized, disruption models considered here; Section~3 describes the observations and the resulting mass, age, and luminosity distributions; and Section~4 compares the \mbox{$M$--$\tau$} properties for clusters in the LMC with the model predictions. Section~5 discusses the physical processes that disrupt star clusters throughout their lives. We summarize our main conclusions in Section~6.  

\section{MODELS}

In FCW09, we derived new analytical formulae for the bivariate mass-age distribution $g(M,\tau)$, defined such that $g(M, \tau) dMd\tau$ is the number of clusters with masses between $M$ and $M+dM$ and ages between $\tau$ and $\tau+ d\tau$, for three idealized models for the disruption of star clusters. We compared predictions from these models with the mass-age distribution of star clusters in the Antennae, and found that 
predictions from one model nicely reproduce the data, while the other two do not. The main goal of this paper is to determine whether or not any of these models describes the star cluster systems in the LMC and SMC. In the first two models, proposed by Boutloukos \& Lamers (2003), clusters are disrupted on a timescale that depends on their masses, with clusters disrupted either suddenly (Model~1) or gradually (Model~2). In the third model, clusters are disrupted suddenly or gradually, on a timescale that is independent of their mass, as indicated by our earlier studies of clusters in
the Antennae (Zhang \& Fall 1999; Fall et~al. 2005, hereafter FCW05; Fall 2006; Whitmore, Chandar, \& Fall 2007, hereafter WCF07; FCW09). For all models, we assume that clusters form with an initial power-law mass function $\psi_0(M_0) \propto M_0^{\beta}$ at a constant rate.\footnote{Our assumption that the cluster formation rate in the LMC and SMC is nearly constant when averaged over large portions of the galaxies is supported by analysis of multi-band imaging of millions of field stars, showing that there have been only minor variations (at approximately the factor of two level) in the rate of star formation over the last few Gyr (Harris \& Zaritsky 2004; 2009). Variations at this level have a negligible impact on the results presented in this work.} Here, we briefly summarize the three disruption models and relevant formulae from FCW09.

\begin{itemize}
\item {\em Model~1. Sudden Mass-Dependent Disruption.}  
This model assumes that clusters retain all of their initial mass until they are destroyed suddenly at an age $\tau_d(M_0)=\tau_{*}(M_0/M_{*})^{k}$, where $k$ is the disruption index, and $\tau_*$ is the characteristic time it takes to disrupt an $M_*=10^4\,M_{\odot}$ cluster (Boutloukos \& Lamers 2003).  The relevant equation from FCW09 for $g(M,\tau)$ is Equation~(6).
It has been claimed that Model~1 provides good fits to the mass and age distributions of clusters in both the LMC (de Grijs \& Anders 2006) and the SMC (Boutloukos \& Lamers 2003; Lamers et~al.\ 2005), with the same parameters $k=0.6$ and $\tau_{*}=8\times10^9$~yr for both galaxies.
We test these claims directly in Section~4.

\item {\em Model~2.  Gradual Mass-Dependent Disruption.}
In this model, we assume that the mass $M$ of a cluster evolves with time $\tau$ according to the equations 
\begin{equation}
dM/d\tau = - M/\tau_d(M),
\end{equation}
\begin{equation}
\tau_d(M) = \tau_* (M/M_*)^k,
\end{equation}
where the exponent $k$ and characteristic disruption timescale $\tau_*$ are adjustable parameters, while $M_*=10^4\,M_{\odot}$ is a fiducial mass scale as before. This second model is an analog of the first, but with continuous evolution (as also noted by Boutloukos \& Lamers).  The value $k = 0$ corresponds to a special case of mass-independent disruption, which describes the disruption of clusters by external gravitational shocks (Spitzer 1987). The case $k = 1$, which implies that $M(\tau)$ has a linear dependence on $\tau$, is appropriate for the standard treatment of stellar evaporation driven by two-body relaxation in clusters of constant mean
density, as regulated by the smooth tidal field of their host galaxies (e.g., Spitzer 1987; Fall \& Zhang 2001). Under some circumstances the evaporation rate is modified when stars scattered to zero or positive energies are subsequently scattered back to negative energies before they can
escape from the cluster, a process known as retarded evaporation. Baumgardt \& Makino (2003) find that a linear $M(\tau)$ relation also provides
a good approximation for retarded evaporation. While $k=1$ is therefore appropriate for both standard and retarded evaporation, the coefficient in the relationship, which is related to the timescale $\tau_*$, has different dependencies on the internal density of the clusters in the two cases. See McLaughlin \& Fall (2008; hereafter MF08) for a thorough discussion of this topic.

\item {\em Model~3. Gradual Mass-Independent Disruption.}  
In this case, both the mass and age distributions are approximated by power-laws (i.e., they have no preferred scales), and are independent of one another, as given by
\begin{equation}
g(M,\tau) \propto (M/M_*)^{\beta} (\tau/\tau_*)^{\gamma}.
\end{equation}
Here, $M_*$ and $\tau_*$ are both arbitrary scale factors that merely specify the units in which masses and ages are measured (in contrast to Models~1 and 2, where $\tau_*$ has physical significance). Equation~(3) is valid if the masses of all clusters decline gradually with the same power-law dependence on age. However this interpretation is not unique. Equation~(3) is also valid if clusters are disrupted suddenly with an age-dependent but a mass-independent probability such that the number of surviving clusters
declines as a power-law in age. Intermediate cases, with different combinations of age-dependent masses and survival probabilities are also possible. Moreover, some clusters may lose only part of their mass, while others are completely destroyed. We refer to all of these situations as gradual mass-independent disruption, because they lead to a gradual decline in the number of clusters at each mass with age, at a fractional rate that is independent of mass.
\end{itemize}

One of the simplest and most informative approaches for comparing observations and predictions is to use averages of $g(M,\tau)$ over several adjacent intervals of age and mass. This approach makes it easy to see features that change over time or with mass, as imprinted by mass-dependent
disruption. These averages over age and mass respectively, are:
\begin{equation}
\bar{g}(M) \equiv \frac{1}{(\tau_2 - \tau_1)}
\int_{\tau_1}^{\tau_2} g(M, \tau) d\tau,
\end{equation}
\begin{equation}
\bar{g}(\tau) \equiv \frac{1}{(M_2 - M_1)}
\int_{M_1}^{M_2} g(M, \tau) dM.
\end{equation}
We have performed the analytical integrations for $\bar{g}(M)$ and $\bar{g}(\tau)$ for each of the three models described above, and give the formulae in Appendix~B of FCW09. This approach allows us to focus on the large-scale shapes of the distributions, rather than relying on a statistical method (e.g., $\chi^2$ minimization) that can easily be overwhelmed by relatively minor systematic errors that result from the dating procedure, such as those which arise during the red supergiant phase (see Section~3.1).
These $\bar{g}(M)$ and $\bar{g}(\tau)$ functions will be our main tool for comparing observations and predictions in Section~4.

\section{OBSERVATIONS}

\subsection{Data and Determination of Cluster Masses and Ages}

We use the catalogs and integrated photometry published by Hunter et~al.\ (2003) to estimate the masses and ages of star clusters in the Magellanic Clouds.  Hunter et~al.\ carefully inspected candidate star clusters (listed in previous surveys) on ground-based \textit{UBVR} images taken at the Michigan Curtis Schmidt telescope at CTIO (Massey 2002), which provide large (but partial) coverage of the LMC ($11~\mbox{kpc}^{2}$) and the SMC ($8.3~\mbox{kpc}^{2}$). They retained objects that could be visually distinguished from the surrounding stellar field and resolved with respect to an isolated star, resulting in a total of 854 clusters in the LMC and 239 clusters in the SMC. These are the most extensive and photometrically homogeneous catalogs currently available, and thus a good dataset to use for a comprehensive study of the cluster populations in the Magellanic Clouds. 
Hunter et~al.\ did not, however, quantify any biases that result from incompleteness in their sample. Therefore in Section~3.3 we perform several tests to assess whether or not biases in the sample significantly impacts our results.

Hunter et~al.\ measured the integrated light for each cluster within an aperture selected to minimize contamination from foreground/background stars, finding good agreement with the integrated colors published in other
catalogs. They did not correct for cluster light beyond the selected aperture.
We find that the $V$-band magnitudes measured by Hunter et~al. are fainter by $0.6$~mag on average, when compared with the total $V$-band magnitudes determined from King profile fits for a subset of clusters (McLaughlin \& van der Marel 2005), with an RMS scatter of $\sim0.5$~mag between the two works. This lack of aperture corrections affects the mass 
estimates of the clusters, and is discussed further below.

We estimate the age $\tau$ and extinction $A_V$ for each cluster by performing a $\chi^2$ fit comparing observed magnitudes with predictions from Bruzual \& Charlot (2003) stellar population models with metallicity $Z=0.008$ for the LMC ($Z=0.004$ for the SMC), a Salpeter (1955) initial mass
function (IMF), and a Galactic-type extinction law (Fitzpatrick 1999).  The best-fit values of $\tau$ and $A_V$ are those that minimize
\begin{equation}
\chi^2(\tau,A_V) = \sum_{\lambda}
W_{\lambda}~(m_{\lambda}^{\mbox{obs}}
- m_{\lambda}^{\mbox{mod}})^2,
\end{equation}
where $m_{\lambda}^{\mbox{obs}}$ and $m_{\lambda}^{\mbox{mod}}$ are the observed and model magnitudes respectively, and the sum runs over
all four bands, $\lambda=U,B,V,R$. The weight factors in the formula for
$\chi^2$ are taken to be $W_{\lambda} = [\sigma_{\lambda}^2 + 
(0.05)^2]^{-1}$, where $\sigma_{\lambda}$ is the photometric uncertainty
determined by Hunter et~al.\ for each band. The mass of each cluster is estimated from the observed $V$-band luminosity (corrected for extinction but not for aperture) and the mass-to-light ratios ($M/L_V$) at the fitted $\tau$ predicted by the models, assuming a distance modulus $\Delta(m-M)=18.50$ for the LMC (Alves 2004), and $18.89$ for the SMC (Harries 
et~al.\ 2003).  We used the same procedure to estimate the ages and masses of clusters in the Antennae galaxies (FCW05 and FCW09), except that for the Magellanic Cloud clusters we have $R$, rather than $I$-band photometry, and no $H\alpha$ measurements.

We have assumed a Salpeter rather than the (more modern) Chabrier stellar IMF mainly to facilitate comparison with our study of clusters in the Antennae, where we made a similar assumption. Another approach would be to assume a Chabrier IMF and to correct the luminosities for the average (0.6~mag) offset found above. We have repeated our dating analysis using this second approach, and find that all the mass and age estimates remain virtually the same. The reason for this can be understood from Figure~1a, which compares predictions from the Bruzual \& Charlot (2003) models 
for Salpeter and Chabrier IMFs, and shows that the adopted IMF makes little
difference in the predicted colors, and hence ages, of the clusters. 
Figure~1b shows that the $M/L_V$, and hence the masses, are reduced by a near constant (age-independent) 40\% for a Chabrier IMF relative to the Salpeter IMF; this reduction in the predicted $M/L_V$ is approximately compensated for by the increased luminosity of the clusters due to the aperture correction. For the Antennae clusters, we made self-consistent aperture corrections based on our $HST$ observations (a procedure not possible for the LMC and SMC clusters), and also adopted the Salpeter IMF. The resulting offset in the mass scale between the Antennae and the Magellanic Clouds is $\approx0.6$~mag ($\approx40$\%) or $\Delta\mbox{log}~M\approx0.24$ (on average). The ages however, are on the same scale, since they are based on the same IMF, and aperture corrections do not affect the colors.

We tested our method by also estimating ages and masses from our $\chi^2$ analysis for different assumptions regarding extinction (using the Calzetti 
et~al.\ 1994 obscuration curve, assuming foreground extinction only, and assuming no extinction) and different sets of stellar population models (Bruzual \& Charlot and GALEV). We discuss the impact that these  assumptions have on the mass and age distributions in Section~3.2. Hunter et~al.\ (2003) also provide an independent estimate of cluster ages, which they determined by comparing their integrated \textit{UBVR} measurements with predictions from the Starburst99 stellar population models (Leitherer et~al.\ 1999) for young ($\tau\leq10^9$~yr) clusters, and with predictions from Searle, Sargent, \& Bagnuolo (1973) and Reed (1985) for older clusters ($\tau > 10^9$~yr). Hunter et~al.\ did not provide mass estimates, so we used the $M/L_V$ predicted by the Bruzual \& Charlot models for the age determined by Hunter et~al.\ for each cluster, and then calculated the mass in the same manner as in our analysis (we refer to these as the ``Hunter et~al.\ masses'' for brevity). In Section~3.2 we repeat our analysis with these independent age and mass estimates, which gives an indication of how sensitive the results are to different assumptions, dating techniques, and models.

We estimate the accuracy of our age determinations as follows. We find that
$\sim$60\% of the clusters have good fits with $\chi^2\leq1$; slightly more than 80\% of the clusters have $\chi^2\leq3$.  This results in typical 1$\sigma$ (internal) uncertainties of $\sim$0.3 in $\log\tau$, corresponding to a factor of $\sim$2 in $\tau$. These internal uncertainties agree well with the external uncertainties of $\approx0.3$--0.4, corresponding to a factor of 2.0--2.5 in $\tau$, based on a comparison with ages determined from absorption-line strengths or from main sequence turnoff fitting for $\approx50$ clusters (e.g., Geisler et~al.\ 1997; Santos et~al.\ 2006; Mackey \& Gilmore 2003a,b; Kerber et~al.\ 2007; Nota et~al.\ 2006; Sirianni et~al.\ 2002). A similar uncertainty was determined for Magellanic Cloud clusters 
by Elson \& Fall (1988) and by de Grijs \& Anders (2006), who also used integrated colors to estimate ages.

We now assess the uncertainties in our mass estimates. The random uncertainties in log~$\tau$ translate to $1\sigma$ uncertainties of $\approx0.3$ in log~$M$, or a factor of two in $M$.  We have already mentioned that the derived masses of the clusters, but not their ages, depend on the assumed IMF in the stellar population models. We find that the mass estimates for the clusters are virtually the same if we adopt a Salpeter IMF
but make no correction for aperture, rather than adopting the Chabrier (2003) IMF, and include a 0.6~mag aperture correction for each cluster.
Similarly, adopting a shorter (longer) distance to the Magellanic Clouds would reduce (increase) the derived masses of the clusters. It is important to note that none of these systematic uncertainties impacts the {\em ratios} of cluster masses or the shape of the mass function presented in Section~3.2.

Figure~2 shows the resulting luminosity-age ($L$--$\tau$) distribution for star clusters in the LMC, and Figure~3 shows the corresponding mass-age ($M$--$\tau$) representation. The solid line in Figure~3 represents $M_V=-4.0$,
corresponding approximately to the limit\footnote{This comes from the change in the mass-to-light ratio of clusters with age, which can be 
approximated by $M/L_V \propto \tau^{0.8}$ for $\tau \ga 10^7$~yr.} $\tau \lea 10^7$ $(M/10^2\,M_{\odot})^{1.3}$~yr and shows that the sample does not contain clusters over the same range of mass at all ages, because clusters fade over time. We use this $M_V=-4.0$ limit to select clusters for the analysis in Section~3.2. Figures~4 and 5 show the corresponding diagrams for clusters in the SMC. Figures~2--5 contain most of the statistical information about clusters in the Magellanic Clouds, and can be projected in different directions to construct the cluster age, mass, and luminosity distributions. When compared with the Antennae, where we restricted our sample to $M_V \lea -9$ (a conservative limit for distinguishing clusters from individual stars), the present study extends the age range by more than an order of magnitude at a fixed mass, and the mass range by more than an order of magnitude at a fixed age.

For some ages, the errors in log~$\tau$ are approximately symmetric and introduce little bias in the age distribution. This is not true, however, in the range $7.0 \lea \mbox{log}(\tau/\mbox{yr}) \lea 7.5$, where the optical
emission from massive clusters is dominated by red supergiant (RSG) stars.
The integrated colors during this phase loop back on themselves, and the fitted ages become degenerate, tending to avoid values $\approx10^7$~yr
and to prefer somewhat higher values. As a result of this type of bias, there likely are features in the unbinned age distribution that are not present in the real age distribution. The relatively empty stripes and similar features in Figures~2--5, while visually prominent, do not affect our conclusions.  For comparison, the bottom panel of Figure~3 shows that cluster ages in the Hunter et~al.\ analysis also bunch up at a few specific ages (which are different from ours), with almost no clusters having derived ages $\tau \lea 3\times10^6$~yr. These types of small-scale features occur regardless of the specific dating analysis that is used, and can vary in location and prominence from one technique to another. We deal with this issue simply by binning on scales over which we believe that any apparent features are real,
$\Delta \mbox{log} \tau \sim 0.6$--0.8.

\subsection{Cluster Age, Mass, and Luminosity Distributions}

All basic results of this paper can be discerned, at least qualitatively, in the \mbox{$\log L$--$\log\tau$} and $\log M$--$\log\tau$ diagrams. When we look down the vertical axis in Figures~3 and 5, we see that the density of clusters increases steadily with decreasing mass. This indicates that clusters have an approximate power-law distribution of masses $\psi(M) \propto M^{\beta}$, which must be steeper than $M^{-1}$, with no turnover or other obvious feature. If we look along the horizontal direction, we see that the number of clusters is approximately even or increases only gradually in equal bins of $\log\tau$ above a given mass, without piling up or dropping off sharply at older ages. This means that the distribution of cluster ages can be described approximately as $\chi(\tau) \propto \tau^{\gamma}$, with $\gamma\approx-1$.  We give a more quantitative treatment below.

The mass function $\psi(M)$, the integral of $g(M,\tau)$ over $\tau$, provides important information about the dynamical evolution of the star clusters. It is obtained by projecting diagonally along the stellar population tracks in Figures~2 and 4 or equivalently projecting horizontally in Figures~3 and 5, and counting clusters in different mass bins. We include clusters brighter than $M_V=-4.0$ (LMC) and $M_V=-4.5$ (SMC), although in the $10^8$--$10^9$~yr interval for the LMC we do include one bin that extends slightly below $M_V=-4.0$, since older, lower mass clusters provide strong constraints on the evolution of clusters. The specific $M$--$\tau$ intervals used to construct the mass function of clusters in the LMC are shown in the 
top panel of Figure~6. In the following, we multiply the mass function by a factor $(\tau_2-\tau_1)^{-1}$ to convert it to $\bar{g}(M)$ as defined by Equation~(3). The {\em amplitudes} of the $\bar{g}(M)$ distributions, in
addition to their {\em shapes}, provide critical information on the evolution of the cluster system.

In Figures~7 and 8, we show $\bar{g}(M)$ for star clusters in the LMC (SMC),
using a typical bin width of $0.5$ ($0.4$) in $\log M$ as a compromise between observational uncertainties in the masses and adequately sampling the mass function. We find that over the plotted mass-age domain, {\em the mass function declines monotonically with increasing mass, with no obvious breaks or bends}, and can be described by a power law\footnote{The mass functions shown in Figures~7 and 8 can be described by power laws up to $M_\mathrm{max}\approx3\times10^5\,M_{\odot}$ in the LMC and to $M_\mathrm{max}\approx10^5\,M_{\odot}$ in the SMC. The expected number of clusters more massive than $M_\mathrm{max}$ is $\approx3$--4 in the LMC (Chandar et~al.\ 2010) and $\approx1$--2 in the SMC, based on extrapolations of these power laws. Thus, we cannot make any definite statements about whether there are or are not physical cutoffs in the mass functions above $M_\mathrm{max}$ in the Magellanic Clouds.}, $\bar{g}(M) \propto M^{\beta}$. The distributions have approximately the same shape for all plotted ages, and are therefore roughly independent of age up to $\tau \approx10^9$~yr. We find $\beta \approx-1.8$ (our masses) and $\beta\approx-1.7$ (Hunter et~al.\ masses) for clusters in the LMC, and $\beta\approx-1.8$ (our masses) and $\beta\approx-2.0$ (Hunter et~al.\ masses) for the SMC. These values of $\beta$ are based on least square fits of the form: $\log\bar{g}(M)=\beta\log M+{\rm const}$.  Any incompleteness in the cluster sample at lower masses would flatten the observed $\bar{g}(M)$ distribution relative to its true shape. We find, however, that $\beta$ hardly changes if we exclude the one or two lowest mass bins from the fit, suggesting that any incompleteness at lower masses does not strongly affect our results. An assessment of how incompleteness impacts the shape of the mass function is discussed more thoroughly in Section~3.3. Based on our extensive experiments with different stellar population models, assumptions about extinction, and binning (both variable and constant), as well as our analysis of the independent age determinations from Hunter et~al., we believe that realistic uncertainties in the exponent $\beta$ are of the order $\Delta\beta \approx 0.2$, and that therefore $\bar{g}(M) \propto M^{\beta}$ with $\beta=-1.8\pm0.2$ describes clusters with ages $\tau \lea 10^9$~yr in both the LMC and SMC.

The age distribution $\chi(\tau)$, the integral of $g(M,\tau)$ over $M$, 
provides important information about the disruption of the clusters. We multiply the age distribution by a factor $(M_2-M_1)^{-1}$ to convert it to $\bar{g}(\tau)$ as defined by Equation~(4). We show $\bar{g}(\tau)$ for star clusters in the LMC (Figure~9) and SMC (Figure~10) based on our age and mass estimates, and also based on the ages determined by Hunter et~al.\ These $\bar{g}(\tau)$ distributions were constructed for {\em mass-limited} samples by counting the number of clusters ($N$) in different age bins, shown as the dashed lines in the bottom panel of Figure~6 for the LMC, and restricted to cover the mass-age domain above the fading lines. The $1\sigma$ uncertainties were taken to be $\pm\sqrt{N}$. In Section~3.1 we estimated $1\sigma$ uncertainties of $\sim0.3$--0.4 in $\log\tau$, and here we use bins twice this wide. Most of {\em the distributions decline starting at very young ages, with no obvious bends or breaks}. Each distribution can be described approximately by a single power law of the form $\bar{g}(\tau) \propto \tau^{\gamma}$. The $\bar{g}(\tau)$ distributions at different masses have the same slope within the uncertainties, and are therefore at least approximately independent of mass. We find typical values of 
$\gamma\approx-0.7$ (our ages) and $\gamma\approx-1.0$ (Hunter et~al.\  ages) for clusters in the LMC, based on simple, linear least-square fits.
These results from two independent dating analyses, plus experiments with different binning and stellar evolution models, suggest $\gamma=-0.8\pm0.2$ for clusters in the LMC. In the SMC, we find $\gamma\approx-0.7$ (our ages) and $\gamma \approx-0.9$ (Hunter et~al.\ ages), where the
youngest bin has been excluded from the fits with the Hunter et~al.\ ages, since these flatten significantly relative to the older data points. These results for clusters in the SMC are similar to those we found previously (Chandar 
et~al.\ 2006) based on the independent cluster sample and age determinations by Rafelski \& Zaritsky (2005).

We have performed extensive experiments with our dating procedure to assess the reliability of these results, as described in Section~3.1. Our experiments suggest that the largest uncertainties in the $\bar{g}(\tau)$ distributions are for the youngest ($\tau \lea 10^7$~yr) clusters, which are most sensitive to the specific assumptions made for extinction when dating the clusters. In general, observations of young clusters in the Magellanic Clouds (e.g., Hunter et~al.\ 1997), and in other galaxies (e.g., M51, Bastian et~al.\ 2005; NGC~5253 and NGC~3077, Harris et~al.\ 2004; Antennae, Whitmore et~al.\ 1999) are affected by extinction in both the Milky Way and in the host galaxy, and it is standard practice to correct for both. However, if we use more extreme assumptions, making {\em no} correction for extinction, or correcting only for extinction in the Milky Way, the youngest data point in the $\chi(\tau)$ distributions move to lower values in both the LMC and SMC, since these assumptions force reddened, young clusters to be assigned ages that are older than their true ages. For ages $\tau \gea10^7$~yr, the $\bar{g}(\tau)$ distributions in the LMC are fairly stable regardless of the detailed assumptions made concerning extinction. This is not true, however, in the SMC, where different assumptions about extinction 
can significantly flatten $\bar{g}(\tau)$ relative to that shown in Figures~9 and 10, with clusters piling up at older ages when no correction for extinction is made. To summarize, we find that the shape of $\bar{g}(\tau)$ is robust for $\tau \gea10^7$~yr in the LMC regardless of the specific assumptions we make, while the results for the SMC are more sensitive to details of our analysis, particularly whether or not we correct for extinction.
However, the $\bar{g}(\tau)$ distributions in both galaxies decline progressively starting at very young ages when standard assumptions are used.

The shape of the luminosity function provides information on the relationship between cluster ages and masses. In the Antennae, we demonstrated that the cluster luminosity and mass functions have the same power-law form (FCW09), even though this is not generally expected for a population of young clusters with a large range in age and hence $M/L$.
Fall (2006) showed analytically that if the cluster age distribution is independent of mass, and the cluster mass function is a power law and independent of age, then the cluster luminosity and mass functions must 
have the same power-law exponents. These conditions are automatically satisfied by $g(M,\tau)$ for Model~3 (but not Models~1 and 2), which provides a good description in the LMC, and possibly in the SMC. In Figure~11, we show that the extinction corrected luminosity functions for clusters in the LMC and SMC can be approximated by a single power law, $\phi(L) \propto L^{\alpha}$, with $\alpha=-1.8\pm0.2$. These have shapes identical to the cluster mass functions shown in Figures~7 and 8, within the uncertainties, as expected from the independence of $M$ and $\tau$, i.e., $g(M,\tau) \propto \psi(M) \chi(\tau)$.

In the Appendix, we show that results from previous works are generally in
good agreement with those presented here for the shape of the mass and age distributions of star clusters in the LMC (although this point is obscured
in the published descriptions of these results). Results from several works in the SMC are also consistent with ours, although there is more disparity for this galaxy. We have determined the shape of the age distribution for  clusters brighter than a given luminosity rather than from mass-limited samples as done here, using the Monte Carlo simulations described in WCF07. Age distributions constructed from luminosity-limited
samples are used below to assess whether or not there are strong biases in either mass or age in the Hunter et~al.\ samples. In the Appendix, we show that the $\bar{g}(\tau)$ distributions presented here for mass-limited samples
are consistent with the age distribution expected from luminosity-limited samples, giving additional confidence that our $\bar{g}(\tau)$ distributions
reflect the actual age distribution of star clusters in the Magellanic Clouds.
We discuss the physical implications of these results for the formation and disruption of clusters in the Magellanic Clouds in Section~5.1.

\subsection{Tests for Biases Due to Incompleteness}

The Hunter et~al.\ samples of Magellanic Cloud clusters  are known to be incomplete in some fashion, since they do not include clusters fainter than
$m_V\approx15$ and do not cover the Magellanic Clouds in their entirety.
What is critical for our work is not that the samples be complete, but that any incompleteness not be a strong function of either mass or age, and hence not significantly affect the shape of the  $\bar{g}(M)$ and $\bar{g}(\tau)$ (mass and age) distributions over the plotted ranges. In other words, we do not require a complete sample, as long as it is unbiased (i.e., representative) with respect to mass and age. Here, we perform three (necessarily) indirect tests on various projections of the $g(M,\tau)$ distribution, which suggest that biases in the sample do not significantly impact the shape of the mass and age distributions presented in Section~3.2.
Several previous studies, described in the Appendix, have used the same dataset to study the formation and disruption of clusters in the Magellanic Clouds, and therefore would be affected by any incompleteness in the same way as in our study.

First, we check whether or not the shape of the mass and age distributions change if we exclude the data points most likely to be affected by incompleteness, namely the lowest mass bin in each $\bar{g}(M)$ distribution, and the oldest age bin in each $\bar{g}(\tau)$ distribution.
Incompleteness would cause $\bar{g}(M)$ and $\bar{g}(\tau)$ in these bins to be lower than their true values. However, we find that they do not deviate significantly from power-law extrapolations to the rest of the data points in the $\bar{g}(M)$ and $\bar{g}(\tau)$ distributions, suggesting that any incompleteness in these bins does not have a strong impact on the shapes of the mass and age distributions.

Next, we check that the $M_V$ limits used here ($M_V=-4.0$  for the LMC and $M_V=-4.5$ for the SMC) are not so faint that we are missing significant numbers of clusters. If the shape of the age distribution constructed from clusters brighter than an $M_V$ limit, i.e., a luminosity-limited sample, changes (flattens) when brighter limits are used, then our $M_V$ limits are too faint. We find that the age distribution barely
changes shape when constructed for clusters at brighter limits of $M_V$ (e.g., $-4.5$, $-5.0$, etc.\ for the LMC), indicating that the limits used in Section~3.2 are not too faint. This result also suggests that the Hunter et~al.\  samples are not limited in a bluer band such as $U$~band rather than
$V$, which could potentially lead to an artificially steep age distribution since clusters fade faster in $U$ than in $V$, with age. In Appendix~A we show that distributions constructed from ages estimated in previous works, but based on different samples of Magellanic Cloud clusters and different limits for $M_V$, are similar to those determined in this work.

Finally, we compare the age distribution constructed in two different ways, by counting clusters above a mass limit and above a luminosity limit. The latter should decline more steeply than the former because some older clusters have faded below the $M_V$ limit. More quantitatively, Monte Carlo simulations show that both distributions can be modeled by power laws, and
that the luminosity-limited distribution has an index that is lower by
$\approx0.4$--0.6. If incompleteness affects the mass-limited but not the luminosity-limited sample, the index for the mass-limited age distribution will be artificially low, and closer to the value determined for the luminosity-limited distribution than the predictions. We find that the age distribution constructed from LMC (SMC) clusters brighter than $M_V=-4.0$ ($M_V=-4.5$) has an index that is always lower by $\approx0.4$--0.5 than the age distribution constructed from the mass-limited samples shown in Figures~9 and 10. Hence the observed and predicted differences between the two distributions are very similar. Based on the results from the three tests
described above, we conclude that the Hunter et~al.\ samples do not have strong biases over the plotted ranges in mass and age,\footnote{The one exception is clusters in the SMC with ages $\tau \lea 10^7$~yr, which likely are under-represented in the Hunter et~al.\ sample; see the Appendix.} and therefore are sufficiently representative of the cluster populations in the Magellanic Clouds for our purposes, i.e., sample incompleteness does not significantly affect the shapes of the $\bar{g}(M)$ and $\bar{g}(\tau)$ distributions presented in Section~3.2. This should be checked in future studies that include artificial cluster experiments during the cluster selection process.

\section{COMPARISON WITH MODELS}

Here, we compare predictions from the three disruption models described in Section~2 with the observed $\bar{g}(M)$ and $\bar{g}(\tau)$ distributions for star clusters in the Magellanic Clouds presented in Section~3. Overall, the $\bar{g}(\tau)$ and $\bar{g}(M)$  distributions for star clusters in the
SMC appear to be similar to those in the LMC, but with poorer statistics.
We therefore only show model comparisons with the LMC, but have checked that all of our conclusions also apply to the SMC, although with larger  uncertainties.

\subsection{Comparison with Model~1: Sudden Mass-Dependent Disruption}

We first consider Model~1, which assumes that clusters are disrupted
suddenly with a power-law dependence on the initial mass, $\tau_d(M_0)=\tau_*(M_0/M_*)^k$, with $M_*=10^4\,M_{\odot}$. We adopt the specific values $k=0.6$ and $\tau_*=8\times10^9$~yr, which have been claimed to give good fits to clusters in the LMC (de Grijs \& Anders 2006) and in the SMC (Boutloukos \& Lamers 2003; Lamers et~al.\ 2005), based on the
indirect methodology developed by Boutloukos \& Lamers. A comparison between the predicted and observed $\bar{g}(M)$ distributions (Figure~12a)
shows that while the {\em shapes} match over the observed range, the {\em amplitudes} do not, because they are predicted to be constant with age but 
are observed to decrease. The situation is even worse for the $\bar{g}(\tau)$ distributions shown in Figure~12b, because the predicted and observed shapes are quite different, with $\bar{g}(\tau)$ predicted to be flat initially,
with a feature determined by the characteristic disruption time, whereas the observations decline as a power law with no obvious feature. {\em We conclude that predictions for the specific combination $k=0.6$ and
$\tau_*=8\times10^9$~yr for Model~1, which has been claimed in several works to give a good description of cluster disruption in the Magellanic Clouds, does not actually match the observed mass and age distributions in either galaxy.}

Why did the Boutloukos \& Lamers approach fail for clusters in the LMC and SMC? The disruption predicted by the combination $k=0.6$, $\tau_*=8\times10^9$~yr is shown as the dashed diagonal line in Figures~3 and 5. According to Model~1, all clusters above this line survive and should be observed, while the region below the dashed line should be empty because the clusters have been destroyed, leading to features in $g(M,\tau)$, $\bar{g}(M)$, and $\bar{g}(\tau)$. However the Hunter et~al.\ data do not reach the masses and ages covered by the dashed line, and therefore these data cannot logically return the claimed values of $\tau_*$ and $k$.  Indeed, de Grijs \& Anders (2006) found no evidence for mass-dependent cluster disruption in the LMC when they plotted the mass function directly (i.e., they
found no bends or breaks). Yet when they applied the method developed by Boutloukos \& Lamers (2003) to the same data, they derived specific values for $k$ and $\tau_*$ that imply access to a mass-age domain for the clusters well beyond the domain that is actually available. The Boutloukos \& Lamers technique is an {\em indirect} one, and assumes that any bend observed in the mass and age distribution is due to mass-dependent disruption. However, bends can appear for other reasons, for example due to artifacts related to systematic errors or to statistical noise, or most problematically when the mass function is constructed from a sample that is limited in luminosity but includes clusters with a wide range of age (as we have shown in FCW09). Thus, the Boutloukos \& Lamers procedure may be more sensitive to selection boundaries in the sample than to any physical relations among the clusters themselves.

\subsection{Comparison with Model~2: Gradual Mass-Dependent Disruption}

Model~2 assumes that clusters are disrupted more gradually than Model~1, 
but still on a timescale that has a power-law dependence on the initial cluster mass. Figure~13 shows that $\bar{g}(M)$ from Model~2 with $\tau_*=8\times10^9$~yr and $k=0.62$ is very similar to $\bar{g}(M)$ from Model~1, and does not match the observations. The predicted $\bar{g}(M)$ distributions for $\tau_* < \mbox{few}\times10^9$~yr with $k=0.62$
flatten earlier (i.e., at a higher mass) than in Figure~12a, causing mismatches with the observations in both shape and amplitude. Higher values of $k$ lead to faster evolution (stronger curvature) in the predicted $\bar{g}(M)$ distributions, while lower values of $k$ lead to slower evolution (weaker curvature). Because the observed mass function for star clusters in the LMC has the same shape but different amplitude for different ages, and the observed age distribution has the same shape but different amplitudes for different masses, no combination of $\tau_*$ and $k>0$ can reproduce the observations. These statements are also true for the mass and age distributions of star clusters in the SMC, as seen from Figures~8 and 11.
A critical point here is that mass-dependent disruption, whether of the power-law form adopted in Models~1 and 2 or any other form, always imprints features in the $\bar{g}(M)$ distribution related to the characteristic disruption timescale. Unless such a feature is observed, there is no evidence that mass-dependent disruption affects a cluster system. {\em We conclude that the disruption of clusters in the Magellanic Clouds has little or no dependence on mass, at least over the $M$--$\tau$ domain studied here.} This conclusion is not affected by the formation history of the clusters, and holds whether clusters form at a constant or variable rate, as long as they form with the same initial mass function (see also FCW09, especially their footnote~7, for further discussion of this point).

Model~2, with $k=0$, is an example of mass-{\em in}dependent disruption.
This preserves the power-law shape of $\bar{g}(M)$, but not $\bar{g}(\tau)$.
Predictions for $\bar{g}(M)$ distributions in the case $k=0$ are shown for
two values of $\tau_*$: $\tau_*=1\times10^8$~yr (Figure~14a) and $\tau_*=2\times10^7$~yr (Figure~14b). While the predicted shape for $\bar{g}(M)$ with $k=0$ matches the observations, these figures show that no single
value of $\tau_*$ gives the correct amplitudes for clusters at different ages in the LMC, with the predictions dropping off faster at older ages than the observations. The reason for this mismatch in amplitudes can be understood from Figure~15, which shows the $\bar{g}(\tau)$ predictions for $k=0$ and several values of $\tau_*$, including the two shown in Figures~14a and 14b.
The predicted distributions start off flat, but change shape at a characteristic age, and then drop exponentially (i.e., faster than any power law). The observed $\bar{g}(\tau)$ distributions on the other hand, are essentially featureless power laws. Therefore, no combination of $k$ and $\tau_*$ can reproduce the observations, and Model~2, like Model~1, must be rejected for clusters in the Magellanic Clouds.

\subsection{Comparison with Model~3: Gradual Mass-Independent Disruption}

We now compare the observations with predictions from Model~3, which posits that the population of clusters is disrupted gradually at a rate that is independent of their mass (although as noted in Section~2, individual clusters may be disrupted gradually or rapidly). Model~3 has $\bar{g}(M) \propto M^{\beta}$ and $\bar{g}(\tau) \propto \tau^{\gamma}$, a direct consequence of the bivariate mass-age distribution $g(M,\tau) \propto M^{\beta} \tau^{\gamma}$. Recall that this model gives a good, first-order description of young clusters in the Antennae (FCW09). In Figure~16a, we compare the observed and predicted $\bar{g}(M)$ distributions in the LMC,
where we have assumed $\gamma=-0.8$ based on the results from
Section~3.2. There is a good match in both shape and in amplitude. The orthogonal projection $\bar{g}(\tau)$ in Figure~16b also shows excellent agreement between predictions and observations. {\em Model~3, with
$\beta=-1.8\pm0.2$ and $\gamma=-0.8\pm0.2$, provides a good, first-order description of the observed mass--age distribution for star clusters in the LMC. } We discuss the physical implications of this result in Section~5.1.

As we have just shown, the joint distribution of masses and ages $g(M,\tau)$ for clusters in the LMC is similar to that found for the Antennae. A nice feature of the present work is that the distributions in the Magellanic Clouds 
extend to clusters with lower masses and to older ages than can be currently distinguished from stars in the Antennae. We show this graphically in Figures~17a and 17b, where we plot $\bar{g}(M)$ and $\bar{g}(\tau)$ distributions for the Antennae, LMC, and SMC, showing that these are all
essentially parallel, and hence that Model~3 also describes---at least approximately---the clusters in the SMC. While studies of young clusters in the solar neighborhood are not as extensive,  the results compiled by Lada \& Lada (2003) support mass and age distributions of the form $\psi(M) \propto M^{-2}$ and $\chi(\tau) \propto \tau^{-1}$, for clusters that have masses lower by an order of magnitude than those studied here.\footnote{Figure~2 of
Lada \& Lada (2003) shows $MdN/d\log M\approx$\,const for embedded (i.e., very young) clusters with $50\,M_{\odot} \lea M \lea 1000\,M_{\odot}$. Figure~3 of Lada \& Lada (2003) shows $dN/d\log\tau \approx$\,const for embedded and non-embedded clusters with $10^6$yr$\lea \tau \lea 10^8$~yr. These results are equivalent to $\psi(M) \propto M^{-2}$ and $\chi(\tau) \propto \tau^{-1}$.} Taken together, {\em these results for clusters in the Antennae, solar neighborhood, LMC, and SMC support the view that there are similar patterns in the $g(M,\tau)$ distributions of young star cluster systems in different, well-studied galaxies.}

\section{INTERPRETATION}

\subsection{Formation and Early Evolution of Star Clusters}

We now seek to understand the results for the LMC, the SMC, and the Antennae in terms of the physical processes that operated during the formation and early evolution of the clusters. This section follows the more extensive discussion in FCW09. The LMC is particularly interesting because its populations of molecular clouds and star clusters have been studied extensively. The mass function of both kinds of objects are approximate power laws, with $\beta\approx-1.7$ for clouds (Rosolowsky 2005; Fukui et~al.\ 2008), and $\beta\approx-1.8$ for clusters (this paper). Thus, the complex processes of converting giant molecular clouds into young star clusters apparently happens in a manner that preserves the shape of the mass function. This suggests that the average efficiency of star formation in
protoclusters---the ratio of the final stellar mass to the initial interstellar mass---is approximately independent of the mass of the protoclusters.

Following their formation, several physical processes are responsible for disrupting star clusters. The dominant processes, which likely operate in the following, approximate sequence and timescales are:
(1)~removal of ISM by stellar feedback, $\tau \la 10^7$~yr; 
(2)~continued stellar mass loss, $10^7~ {\rm yr} \la \tau \la 10^8~ {\rm yr}$;
(3)~tidal disturbances by passing molecular clouds, $\tau \ga 10^8$~yr; and
(4)~escape of stars driven by internal two-body relaxation (hereafter referred to as ``evaporation''), for clusters with $M \lea M_p$, as given by Equation~(7) below. We discuss the first three processes here, and return to the last process in the next subsection.

One of the earliest processes responsible for the disruption of star clusters is
the removal of left-over ISM by the feedback activity of massive stars (e.g., photoionization, radiation pressure, stellar winds, and supernovae). An important consequence of this rapid mass loss is that many protoclusters cannot remain gravitationally bound (e.g., Hills 1980). Afterwards, clusters continue to lose mass due to stellar evolution, and will be depleted by $\sim$40\% over $\sim\mbox{few}\times10^8$~yr, with most of this mass lost in the first $\mbox{few}\times10^7$~yr. This continued mass loss can unbind fragile clusters in a tidal field. Both of these disruption processes are internally driven, and will occur regardless of the specific properties of the host galaxy. Therefore, we would expect these early disruption processes to affect young clusters in different galaxies in a similar way. This expectation is supported by observations of clusters in the solar neighborhood, the merging Antennae galaxies, and the LMC and SMC, which all have a similar shape for $\bar{g}(\tau)$ (as shown in Figure~17 and discussed in Section~4).

It has been suggested, on theoretical grounds, that the removal of the ISM strongly affects the shape of the cluster mass function by disproportionately disrupting lower mass clusters. Some models predict that an initial power-law mass function with $\beta=-2$ will evolve quickly after removal of the ISM, on timescales $\tau \lea 10^7$~yr, resulting in a flatter mass function with $\beta\approx-1$ (see Figure~4 in Baumgardt et~al.\ 
2008),\footnote{Baumgardt et~al.\ (2008) state that gas expulsion will introduce a peak in the mass function near $M_p\approx10^5\,M_{\odot}$.  However their Figure~4 shows a different result. This figure suggests that while the predicted mass function is significantly flatter after gas expulsion, a peak only occurs on much longer timescales, after other disruption processes, particularly relaxation-driven stellar evaporation, have had a chance
to further erode the cluster system.} or one that has a peak near $10^5\,M_{\odot}$ (e.g.,  Kroupa \& Boily 2002; Parmentier et~al.\ 2008). However, these predictions are ruled out by observations. We see from Figures~7 and 8 that the mass functions of clusters in the Magellanic Clouds with ages $\tau < 10^7$~yr and $\tau > 10^7$~yr have essentially the same shape, 
indicating that early disruption processes do not depend strongly on the masses of the clusters. This result is explained nicely by models in which the stellar feedback is momentum-driven rather than energy-driven (Fall et~al.\  2010).

Clusters are also subject to tidal disturbances by passing molecular clouds (Spitzer 1958; Binney \& Tremaine 2008). The timescale for this process depends on the number and density of molecular clouds in a galaxy. For example, Binney \& Tremaine (2008) estimate a disruption time of $\tau_d\approx3\times10^8$~yr for clusters in the solar neighborhood, due to interactions with molecular clouds. If conditions in the Magellanic Clouds are similar, this estimate of $\tau_d$ might also apply there. We argued in FCW09 that interactions with molecular clouds would have little or no effect
on the shape of the mass function of clusters. In contrast, Gieles et~al.\  (2006) claim that such interactions would preferentially disrupt low-mass clusters. We see no evidence for this effect, either in the Antennae or in the Magellanic Clouds over the ranges of masses and ages we have examined.
We discuss the physical reasons for and implications of this result in a separate paper  (S.~M.\ Fall \& R.~Chandar, in prep).

It is likely that the rate of disruption is somewhat different for each of 
the processes discussed above, and that with perfect data one might observe
features marking the transition between them. However, it is also likely that the combination of processes, plus the errors in the masses and ages caused by imperfect observations, have washed out such features, leaving a relatively smooth age distribution. For these reasons, our power-law model, 
$\bar{g}(\tau) \propto \tau^{\gamma}$ is likely to be a simple  approximation to a complex situation including several different physical processes which predominate at different times, rather than an exact description of a single process working in isolation.

For the Antennae, we argued that $\bar{g}(\tau)$ {\em primarily} reflects
the disruption rather than the formation of clusters,  because it has the same sharp peak at $\tau \lea 10^7$~yr in several large regions inside the Antennae galaxies separated by distances of order 10~kpc. There are no hydrodynamical processes that could synchronize a burst of cluster formation this precisely over such large separations (FCW05; WCF07). In the present work, we found that star clusters in the LMC and SMC, two galaxies which are currently in a more quiescent phase than the Antennae, nevertheless have similar age distributions. A recent analysis of the star clusters in the (closest) collisional ring galaxy NGC~922, also shows a similar form for the age distribution (Pellerin et~al.\ 2009). It is far more likely that the clusters in all of these very different galaxies have similar disruption histories, than it is that they have similar formation histories and that we also happen to be observing them all at the same special time when the rate of formation is just now peaking.

An alternative explanation, that the declining shape of $\bar{g}(\tau)$ in the LMC over the last few Gyr is due to a large increase (by a factor $\approx100$) in the rate at which clusters formed from the past to the present, was suggested by Parmentier \& de Grijs (2008). In addition to the problem discussed above, another powerful argument against this suggestion is that the star formation history of the SMC and LMC determined from millions of individual stars is nearly constant over this timescale, with variations of only a factor of two or less (Harris \& Zaritsky 2004, 2009). Nevertheless, we expect variations in the formation rate of clusters to have some effect on their age distribution, as reflected in minor differences in the exponent $\gamma$ among different galaxies, likely at the level $\Delta\gamma \approx 0.2$. This is consistent with our finding $\gamma = -1.0 \pm 0.2$ for the Antennae and $\gamma = -0.8 \pm 0.2$ for the LMC and SMC.

\subsection{Late Evolution of Star Clusters}

The most important long-term disruptive process for star clusters is the
gradual escape of stars driven by internal two-body relaxation (e.g., Fall \& Zhang 2001; Prieto \& Gnedin 2008). This process depletes the mass of each cluster approximately linearly with time, $M(\tau) \approx M_0 - \mu_\mathrm{ev}\tau$, equivalent to $k=1$ in Model~2, at a rate $\mu_\mathrm{ev}$ that depends primarily on the mean internal density of the cluster. The evaporation rate  $\mu_\mathrm{ev}$ has a slightly different
dependence on density for tidally limited clusters in the cases of standard and retarded evaporation, with $\mu_\mathrm{ev} \propto \rho_t^{1/2}$
(with $\rho_t=3M/4\pi r_t^3$) for standard evaporation, and $\mu_\mathrm{ev} \propto \Sigma_t^{3/4}$ (with $\Sigma_t = M/\pi r_t^2$) for retarded evaporation. See McLaughlin \& Fall (2008) for a detailed discussion of these
dependencies. Elson et~al.\ (1987) found that the density profiles of young clusters in the LMC (with $\tau \sim 10^8$~yr) extend out to and even beyond their tidal radii (sometimes also referred to as Jacobi radii) as determined from a detailed study of the tidal field of the LMC (see also Lupton
et~al.\ 1989). Thus, these clusters already fill their Roche lobes, and satisfy the main assumption underlying the simple evaporation model of McLaughlin \& Fall.

For a population of clusters with a single age, which is effectively true for ancient globular clusters (GC),  a peak or turnover in the mass function at $M_p$ results from the higher fractional mass loss rates ($d\ln M/d\tau$) for low-mass clusters than for high-mass clusters. This peak should increase
with the internal density of the clusters, since stars evaporate more quickly from denser clusters, leading to their earlier disruption relative to lower density clusters. MF08 found this exact trend of $M_p$ with $\rho_h$ in the Milky Way globular cluster system, and presented a simple evaporation model with $\mu_\mathrm{ev} \propto \rho_h^{1/2}$, where $\rho_h$ is the internal half-mass density (a good approximation to both $\mu_\mathrm{ev} \propto \rho_t^{1/2}$ and $\mu_\mathrm{ev} \propto \Sigma_t^{3/4}$ in practice). Putting in the coefficients, the peak mass\footnote{We note that the mass function strictly only shows a peak when it is plotted as $dN/d\log M$.  In this work, we have plotted $dN/dM$ throughout.  In this form, the ``peak'' of the mass function $M_p$ corresponds to the point where the distribution flattens, or has a ``knee.''}  is given by 
\begin{equation}
M_p = 3.1 \times (\tau / 13\times10^9~\mathrm{yr}) (\rho_h / 10^3)^{1/2} 10^5\,M_{\odot} .
\end{equation}
Assuming the median half-mass density for GCs in the Milky Way of $\hat{\rho_h}=246\,M_{\odot}~\mbox{pc}^{-3}$ (MF08), at an age $\tau=13$~Gyr, Equation~(7) reproduces the observed peak mass $M_p=1.6\times10^{5}\,M_{\odot}$ for the Galactic GC system. The same equation, with a similar coefficient, also explains the observations of GCs in the Sombrero galaxy (Chandar, Fall, \& McLaughlin 2007).

Here, we apply the MF08 model to clusters in the LMC for the first time. The LMC has 15 known ancient GCs, which reside in a rotating disk with their younger counterparts (e.g., Freeman et~al.\ 1983). We estimate their masses from the total $V$-band luminosities determined by McLaughlin \& van der Marel (2005) from King (1966) model fits to their radial profiles,
and assume a typical $M/L_V=1.5$, the median value found for the Milky Way GC system. This gives a median mass of $\hat{M} \approx 2\times10^{5}\,M_{\odot}$ for the GC system in the LMC, very similar to that found for the GC systems in other galaxies such as the Milky Way and the Sombrero.
We compute the internal half-mass density of clusters from $\rho_h = (M/2)/(4\pi r_h^3/3)$, where $r_h$, the 3D half-mass radius  is determined from the 2D half-mass radius $r_\mathrm{hp}$ and the standard 3D--2D conversion $r_h = (4/3) r_\mathrm{hp}$ (Spitzer 1987). For the sample of 15 globular clusters in the LMC, we find a median half-mass density $\hat{\rho_h}=295\,M_{\odot}~\mbox{pc}^{-3}$, similar to the median density of 
$246\,M_{\odot}~\mbox{pc}^{-3}$ found by MF08 for the Galactic globular cluster system. Globular clusters in the LMC have ages indistinguishable from their Galactic counterparts (e.g., Brocato et~al.\ 1996), and for an assumed age $\tau=13$~Gyr and the observed value of $\hat{\rho_h}$, Equation~(7) gives $M_p \approx 1.7\times10^{5}\,M_{\odot}$, very similar to the observed value for $\hat{M}$.

Next, we consider the $10^8$--$10^9$~yr-old LMC clusters, for which we plotted $\bar{g}(M)$ in Figure~7. We find a median internal density of 
$\rho_h\approx20\,M_{\odot}~\mbox{pc}^{-3}$ for the nine clusters in this age interval that have profile fits from McLaughlin \& van der Marel 
(2005).\footnote{While Hunter et~al.\ 2003 provide FWHM measurements for the clusters in their sample, there is a poor correlation between their ground-based size estimates and those measured by McLaughlin \& van der Marel 2005 from high-resolution \emph{HST} images. Therefore we decided not to use the Hunter et~al.\ size estimates in this work.} This is approximately 14 times lower than that found for the ancient LMC clusters.
This difference in density is primarily due to the larger size of the $10^8$--$10^9$~yr-old clusters compared with that for the old GCs ($\hat{r_h}=8.2$~pc vs.\ $\hat{r_h}=3.4$~pc). Assuming that the median density of $\rho_h\approx20\,M_{\odot}~\mbox{pc}^{-3}$ for these nine clusters is representative of the entire $10^8$--$10^9$~yr old population, Equation~(6) predicts $M_p\approx10^3\,M_{\odot}$ for $\tau=3\times10^8$~yr, the median age of clusters in this bin.  A flattening of $\bar{g}(M)$ near this mass would be near the edge and possibly below the plotted range in Figure~7. Therefore, the fact that we do not detect any flattening in the mass function for $10^8$--$10^9$~yr-old clusters in the LMC is consistent with the MF08 model predictions.

\section{SUMMARY AND CONCLUSIONS}

We have compared the mass-age distribution $g(M,\tau)$ of star clusters in the Magellanic Clouds with three idealized disruption models: (1)~sudden mass-dependent disruption, (2)~gradual mass-dependent disruption, and 
(3)~gradual mass-{\em in}dependent disruption, to determine whether any of them describe the properties of star clusters in these relatively quiescent
galaxies. We compared the integrated \textit{UBVR} photometric measurements for 854 LMC and 239 SMC clusters cataloged by Hunter et~al.\ (2003), with predictions from population synthesis models in order to estimate $M$ and $\tau$ for each cluster. We constructed $\bar{g}(M)$ and $\bar{g}(\tau)$ (mass and age) distributions, and tested them for sensitivities with respect to stellar population models, dating techniques, and extinction corrections. These distributions in the LMC are fairly robust, particularly for $\tau \gea10^7$~yr, but in the SMC can flatten significantly given the most extreme (although unrealistic) assumptions. We find that the shape of the cluster mass and age distributions are in reasonably good agreement with the results from most previous studies in the LMC, and with several studies in the SMC (as explained in detail in the Appendix). While there are remaining uncertainties, such as not knowing the precise completeness of the cluster samples or the aperture corrections for individual objects, which affect the results presented here as well as those from previous works, we believe that these will not significantly alter our
conclusions. Our main findings are:

(1) Model~3, with mass-{\em in}dependent disruption provides a good match to the observed mass-age distribution of star clusters in the Magellanic Clouds, at least over the $M$-$\tau$ range studied here: roughly $\tau \lea 10^7(M/10^2\,M_{\odot})^{1.3}$~yr. The model can be expressed as $g(M,\tau) \propto M^{\beta} \tau^{\gamma}$, with best-fit exponents $\beta=-1.8\pm0.2$ and $\gamma=-0.8\pm0.2$.

(2) Models~1 and 2 give poor descriptions of the mass and age distributions of star clusters in the Magellanic Clouds, for all combinations of the adjustable parameters $k$ and $\tau_*$. This contradicts previously published
claims (Boutloukos \& Lamers 2003; Lamers et~al.\ 2005; de Grijs \& Anders 2006) that these models, with the specific parameters $k\approx0.6$ and $\tau_*=8\times10^9$~yr, provide a good description of these cluster systems.

(3) The luminosity function of star clusters in the Magellanic Clouds is well represented by a single power law, $dN/dL \propto L^{\alpha}$, with $\alpha=-1.8\pm0.2$.  Within the uncertainties, the luminosity function has the same power-law exponent as the mass function. Fall (2006) has shown that this is a direct consequence of the fact that the cluster age and mass
distributions are independent of one another.

(4) The young clusters in the LMC, with $\tau \lea 10^9$~yr, have substantially lower internal (half-mass) densities than the old globular clusters
with $\tau \approx10^{10}$~yr. Thus, the young clusters have correspondingly lower rates of evaporation, and this explains the absence of a bend in their mass function, according to the McLaughlin \& Fall (2008) model.

Our main result is that the bivariate mass-age distribution for star clusters in
the LMC and likely in the SMC can be approximated by $g(M,\tau) \propto M^{\beta} \tau^{\gamma}$ with indices $\beta$ and $\gamma$ similar to those we found for star clusters in the Antennae galaxies (i.e., $\beta\approx-2$ and $\gamma\approx-1$; Zhang \& Fall 1999; FCW05; WCF07; FCW09).
The Magellanic Cloud clusters extend the domain of validity for this result to older ages and to lower masses than can currently be studied for clusters in the Antennae. The similarity found for $g(M,\tau)$ provides a strong test that
the processes which dominate the formation and early disruption of star clusters are similar from galaxy to galaxy, because the Antennae and the Magellanic Clouds represent different environments for star and cluster formation: two large, interacting galaxies versus two small, relatively quiescent galaxies.  We hypothesize that this simple formula for $g(M,\tau)$ describes, at least approximately, the early evolution of all star clusters (open, globular, populous, proto-, and super-), in many if not most galaxies.
We discussed the following, approximate sequence for disruption processes 
that shape $g(M,\tau)$ over the first $\approx10^9$~yr:
(1)~removal of the ISM by the activity of massive stars (e.g., winds, radiation, supernovae), $\tau \la 10^7$~yr, 
(2)~continued mass loss due to stellar evolution, $10^7 {\rm yr} \la \tau \la 10^8~{\rm yr}$, and
(3)~tidal disturbances by passing molecular clouds, $\tau \ga 10^8$~yr.
These processes are expected to operate in a fashion that will not change the shape of the mass function with age, consistent with the observations.
If this picture turns out to be generally valid, it will mean that the main difference between populations of young clusters in different galaxies is simply in the normalization (amplitude) of $g(M, \tau)$, and hence in the number of clusters formed.

\acknowledgements{We are extremely grateful to Deidre Hunter and
Bruce Elmegreen for providing their Magellanic Cloud cluster catalogs and photometry, and for helpful discussions about them. We thank Dean McLaughlin, Francois Schweizer, and Bruce Elmegreen for comments on an earlier version of the manuscript, and the referee for helpful suggestions that improved our paper. SMF acknowledges support from the Ambrose Monell Foundation and from NASA grant AR-09539.1-A. }

\appendix

\section{COMPARISON WITH PREVIOUS WORKS}

There have been a number of recent studies of the mass and age distributions
of star clusters in the LMC and SMC, some of which have claimed shapes for $\psi(M)$ and $\chi(\tau)$ different from those presented here. Other studies have found similar shapes but interpreted them differently in terms
of the formation and disruption of clusters. Nevertheless, we show here that there is actually fairly good agreement regarding the shapes of the mass and age distributions, particularly in the LMC. Some of the apparent differences arise from confusion between distributions plotted in $\tau$ versus $\log\tau$, and others because authors have made initial assumptions that have led them to different conclusions. Below, we summarize these previous results and assess how they compare with those presented here.

\textit{LMC}: Elson \& Fall (1985) derived one of the first age distributions for clusters in the LMC, and pointed out that an age distribution constructed for a mass-limited sample gives more information about dynamical processes than that for a luminosity-limited sample, because the latter is strongly affected by the fading of the clusters. Regardless, many works have presented $\chi(\tau)$ based on luminosity-limited samples. We find that $\chi(\tau)$ constructed from a sample of clusters brighter than a given 
$V$-band magnitude limit can be approximated by a power law, $\chi(\tau) \propto \tau^{\gamma_L}$, with $\gamma_L\approx-1.15$, based on our age and mass determinations, and $\gamma_L\approx-1.25$ based on the Hunter et~al.\ age estimates. The $\gamma_L$ index is always steeper by $\approx-0.4$  to $-0.5$ when compared with $\gamma$ measured for a mass-limited sample (recall that we found $\gamma\approx-0.8$).

Girardi et~al.\ (1995) presented ages for $\approx450$ clusters based on a re-calibration of the S-sequence. We find  that the age distribution from this luminosity-limited sample can be approximated by a single power-law,
with $\gamma_L \approx-1.1$. Using a sample $\approx30$\% larger, Bica et~al.\ (1996) tabulate the number of clusters and associations in different SWB-types (regions in a two-color diagram), which correspond to different
ranges in age. We find that the age distribution constructed from this luminosity-limited sample gives $\gamma_L\approx-1.2$ for $\tau \gea 10^7$~yr. Therefore, both the Girardi et~al.\ and Bica et~al.\ analyses result in
power-law age distributions with exponents similar to ours.

Elmegreen \& Efremov (1997) plotted the luminosity function of clusters in
relatively narrow intervals of age, effectively the mass function, and found that it could be approximated by a power law with $\beta\approx-2$ for clusters with ages in the range $10^8~\mbox{yr} \leq \tau \leq 10^9~\mbox{yr}$. For younger clusters, the mass functions are flatter but also noisier,
with $\beta\approx-1.5$ for $10^7~\mbox{yr} \leq \tau \leq 10^8~\mbox{yr}$ and $\beta\approx-1.0$ for $10^6~\mbox{yr} \leq \tau \leq 10^7~\mbox{yr}$. They also found that the age distribution is a declining function for
$\tau\lea10^9$~yr, and suggested that a significant fraction of clusters do not survive to old age. Elmegreen \& Efremov's results and conclusions are qualitatively (although not always quantitatively) similar to those presented here.

Hunter et~al.\ (2003) presented the cluster sample and photometry used in this work. They inferred that the initial cluster mass function could be 
described as a power law, $\psi(M) \propto M^{\beta}$, with $\beta$ between $-2.0$ and $-2.4$, based on a technique that relies on the scaling of
the maximum cluster mass with age (rather than directly from the mass function itself). Their value for $\beta$ is generally in good agreement with, although somewhat steeper than, the index of $\beta \approx-1.7$ that we found directly based on their dating analysis in Section~3.2. These differences in $\beta$ may result from their indirect technique for inferring the shape of the initial mass function, which is based on the assumption that 
the age distribution of the clusters is flat in $\tau$ rather than $\log\tau$ (as we have found here from the same dataset).

De Grijs \& Anders (2006) estimated the masses and ages of star clusters in the Hunter et~al.\ catalog, comparing the data with predictions from the GALEV stellar population models. They determine the mass function in different intervals of age (their Figure~8, which is analogous to our Figure~7),
and find $\beta\approx-1.7$ to $-1.8$ for $\tau \lea 6\times10^7$~yr and 
$\beta\approx-2.0$ for $6\times10^7 < \tau \lea 10^9$~yr. These values of $\beta$ are within the $\pm0.2$ uncertainties that we estimated for $\beta$ in Section~3.2, although de Grijs \& Anders claim that the power-law indices are different for clusters younger and older than $\tau=6\times10^7$~yr based on the formal errors, which are smaller than $\approx0.2$.\footnote{De Grijs \& Anders (2006) plot their mass function in {\em cumulative} mass intervals, restricting the samples with older clusters to higher masses to avoid incompleteness due to clusters fading below the completeness limit, whereas we plot the mass function in differential form in relatively narrow age slices.} Their results can be re-expressed in our notation as $g(M,\tau) \propto M^{-2}\chi(\tau)$, since the shape of the mass function is nearly the same at different ages. In particular, we note that the mass function for
clusters with ages $\log\tau \lea 9.75$~yr shows no deviation from a power law down to the lowest plotted mass, i.e., there is no evidence in the de Grijs \& Anders results for any mass-dependent disruption of star clusters in the LMC. This is  similar to the results presented here. Nevertheless, de Grijs \& Anders applied the {\em indirect} methodology developed by Boutloukos \& Lamers (2003) and found a characteristic disruption timescale of $\tau_*=8\times10^9$~yr and a disruption index $k=0.56$. As shown in Figure~3, however, these values of $\tau_*$ and $k$ imply a $M$--$\tau$ range beyond that observed in the Hunter et~al.\ data.

Parmentier \& de Grijs (2008) noticed an inconsistency between the shape of the age distribution and the value of $\tau_*$ inferred by de Grijs \& Anders (2006), and re-investigated the LMC cluster system, again based on the Hunter et~al.\ data. This new study concludes that there is no direct evidence that lower mass clusters are disrupted faster than higher mass clusters in the LMC. They find that the cluster age distribution declines steadily with age for mass-limited samples covering different mass intervals (although the decline is much shallower for the most massive clusters), 
similar to the results presented here. Parmentier \& de Grijs, however, claim that the physical reason for this decline in $\chi(\tau)$ is a decrease in the rate of cluster {\em formation} by a factor of $\approx100$ from the present to a few Gyr ago, rather than to the disruption of clusters. As discussed in Section~5.1, there are three problems with this interpretation: (1)~it requires that the rate of cluster formation increased by a huge factor over a
relatively short time interval; (2)~the age distribution for young clusters in several galaxies (the SMC, LMC, Antennae, solar neighborhood, NGC~922),
all have very similar shapes, making it much more likely that disruption processes, which are largely independent of the host galaxy over the first $\approx\!10^8$~yr, are responsible for this shape, rather than a fortuitous increase in the rate of cluster formation in all of these galaxies at exactly this time in their history; and (3)~the star formation history of the Magellanic
Clouds determined from individual stars has been nearly constant (to within a factor of 2), over the last few Gyr. Parmentier \& de Grijs (2008) assumed {\em a priori} that mass-independent disruption ended $\approx10^7$~yr ago,
leading them to conclude that it is the formation rather than the disruption of clusters that is primarily responsible for the steep decline in $\chi(\tau)$.

To summarize, essentially all recent studies are consistent with the general results presented here regarding the shape of the mass and age distributions
of star clusters in the LMC, particularly for ages $\tau \gea10^7$~yr. Some authors however, have reached different conclusions regarding the interpretation of this shape in terms of the formation and disruption of clusters.

\textit{SMC}: Just as for the LMC, some previous works have presented age distributions in the SMC based on cluster samples that are limited in luminosity rather than in mass. We find that the age distribution constructed for a $V$-band limited sample gives $\gamma_L\approx-1.2$ from our age determinations, and $\gamma_L\approx-1.3$ from the Hunter et~al.\ ages. Hodge (1987) presented an early age distribution based on an approximate dating technique that compared the brightest star in a cluster with theoretical predictions, and suggested that clusters in the SMC have a longer characteristic disruption time than their counterparts in the Milky Way.

More recently, several works based on the sample and photometry of Rafelski \& Zaritsky (2005) have reached apparently contradictory conclusions.
The main differences come from assumptions made regarding (internal) extinction corrections during the dating procedure. Rafelski \& Zaritsky estimated the ages for 195 clusters in the SMC by comparing integrated \textit{UBVI} measurements with predictions from STARBURST99 and GALEV stellar population models, both including and excluding corrections for 
extinction in the SMC. Their sample avoids emission-line regions and is thus (deliberately) biased against the youngest ($\tau \lea 10^7$~yr) clusters.
Based on {\em extinction-corrected} ages for a sample limited in the $V$ band, they find that the age distribution declines smoothly and can be approximated by a power law. Their Figure~12 shows $\chi(\tau) \propto \tau^{-1.1}$, although they mistakenly quote this as\footnote{We downloaded the Rafelski \& Zaritsky (2005) data and confirm that it gives $dN/d\tau \propto \tau^{-1.1}$, as shown previously in WCF07.} $\chi(\tau) \propto \tau^{-2.1}$. An index $\gamma_L\approx-1.1$ is similar to the results mentioned above, based on analyses of the independent cluster sample of Hunter et al.

Using the same extinction-corrected ages as Rafelski \& Zaritsky, we previously found that the {\em mass}-limited age distribution for star clusters
older than $10^7$~yr can be described by a power-law, with an exponent $\gamma\approx-0.8$ (Chandar, Fall, \& Whitmore 2006). This result is in good agreement with those presented here based on the Hunter et~al.\  sample and photometry.

Gieles, Lamers, \& Portegies Zwart (2007) used the ages from Rafelski \& Zaritsky that were {\em not corrected} for extinction, and found that the age distribution for {\em mass}-limited samples is approximately flat  ($\gamma\approx0$), and for {\em luminosity}-limited samples has a power-law index $\gamma_L\approx-0.7$ to $-0.8$. These are significantly flatter than the mass- and luminosity-limited age distributions determined
by Chandar et~al.\ (2006) and by Rafelski \& Zaritsky (2005).\footnote{To confuse matters further, Gieles et~al.\ (2007) claimed, incorrectly, that the Chandar et~al.\ (2006) analysis was based on luminosity- rather than mass-limited samples.} They are also significantly flatter than the results presented in this work using the sample and photometry of Hunter et~al. The flatter age distribution found by Gieles et~al.\ results from their use of ages estimated without corrections for extinction. We found a similar age distribution in Section~3 when we neglected extinction corrections, with clusters piling up at older ages as in Figure~1 of Gieles et~al. However, the current state of knowledge suggests that this result is not physical: clusters do not pile up at older ages in the SMC, given that only $\approx30$--40 clusters are known to have ages $\tau \gea10^9$~yr (based on the analysis of color-magnitude diagrams; e.g., Piatti et~al.\ 2005a,b, 2007), out of $\approx300$--400 currently known clusters, i.e., only 10\% of the clusters cover $\gea90$\% of the actual time interval.

Chiosi et~al.\ (2006) determined the ages of 475 SMC clusters younger than
$\approx10^9$~yr from isochrone fitting of the main sequence turnoff region.
They present two different age distributions---one that includes all of their objects and another that excludes all objects classified as ``associations.''
For the full sample, we find that their age distribution has $\gamma_L\approx-1.1$, similar to the results presented here. Their distribution constructed from the subsample that excludes objects they classify as associations is essentially flat in $\log\tau$ for ages $\tau \lea 10^8$~yr, and 
declines afterwards. However, in rejecting the objects designated as associations, Chiosi et~al.\ retain only a single SMC cluster younger than $10^7$~yr. Their selection criteria almost certainly introduces an artificial bias, 
since the SMC is known to have formed many clusters in the last $\approx10^7$~yr (for e.g., see Sabbi et~al.\ 2007 for a list of 16 recently discovered
compact clusters with $\tau < 10^7$~yr, located within a very small portion of the SMC. The Sabbi et~al. study indicates that current catalogs of clusters in the SMC are incomplete at these ages.)

De Grijs \& Goodwin (2008) recently performed an independent dating 
analysis of star clusters in the SMC, based on the Hunter et~al.\ sample and observations. They estimate that at most, 30\% of clusters in the SMC disrupt over the first $\approx10^7$~yr. Our experiments in Section~3 showed that the estimated number of clusters with $\tau \lea 10^7$~yr is sensitive to details of the extinction correction. Coupled with the incompleteness of current cluster catalogs in the SMC, we believe that this rate is highly uncertain. The age distributions plotted by de Grijs \& Goodwin (2008) for clusters older than $10^7$~yr, in mass-limited samples, are not flat however, but rather decline with age, quite different from the shape found by Gieles et~al.\ (2007) but similar to that found here.

Overall, the shape of the age distribution found previously for clusters in the SMC is in reasonably good agreement with that presented here, particularly in the range $\tau=10^7$--$10^9$~yr. This is consistent with our results for clusters in the LMC and in the Antennae. We caution, however, that results for the SMC are less robust because they are sensitive to several assumptions, particularly to the adopted extinction correction. There is additional uncertainty for $\tau\lea10^7$~yr clusters because current catalogs are still incomplete for these ages, making it difficult to establish the fraction of clusters in the SMC that dissolve during the earliest phases of mass loss.

\clearpage

\begin{figure}
\plottwo{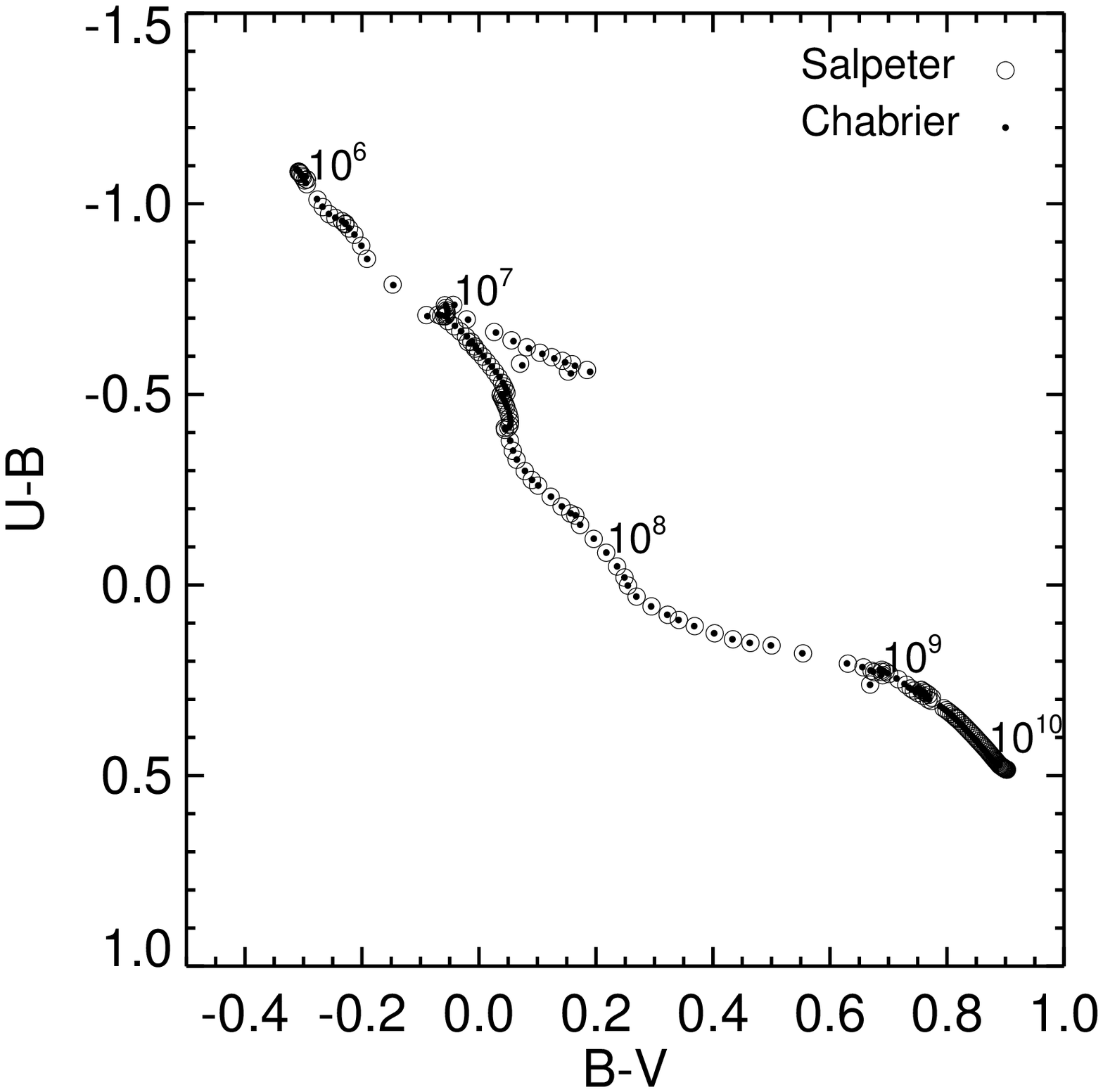}{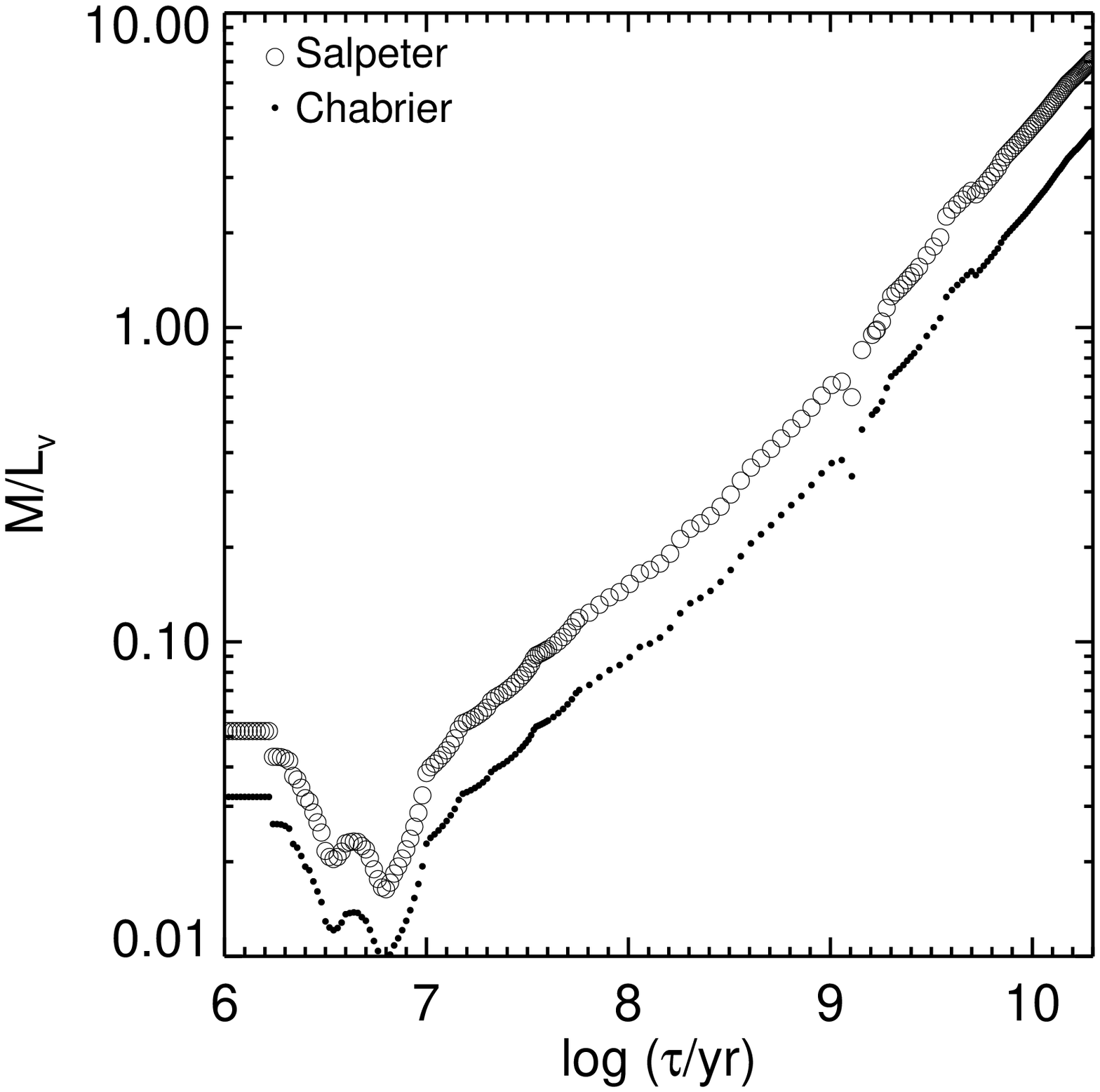}
\caption{Predictions from the stellar population models of Bruzual \& Charlot (2003) with metallicity $Z=0.008$, for an assumed Salpeter IMF (large, open circles) and a Chabrier IMF (small, filled circles). The $U-B$ vs.\ $V-I$ two-color diagram on the left shows that the adopted IMF has little impact
on the predicted colors of star clusters at all ages (i.e., all of the small, filled circles fall within the larger open circles). The $M/L_V$ ratio predicted for a Chabrier IMF is lower than for a Salpeter IMF by a nearly constant value
($\approx40$\%) at all ages, as shown on the right.
}
\label{fig:sspcomp}
\end{figure}

\begin{figure}
\plotone{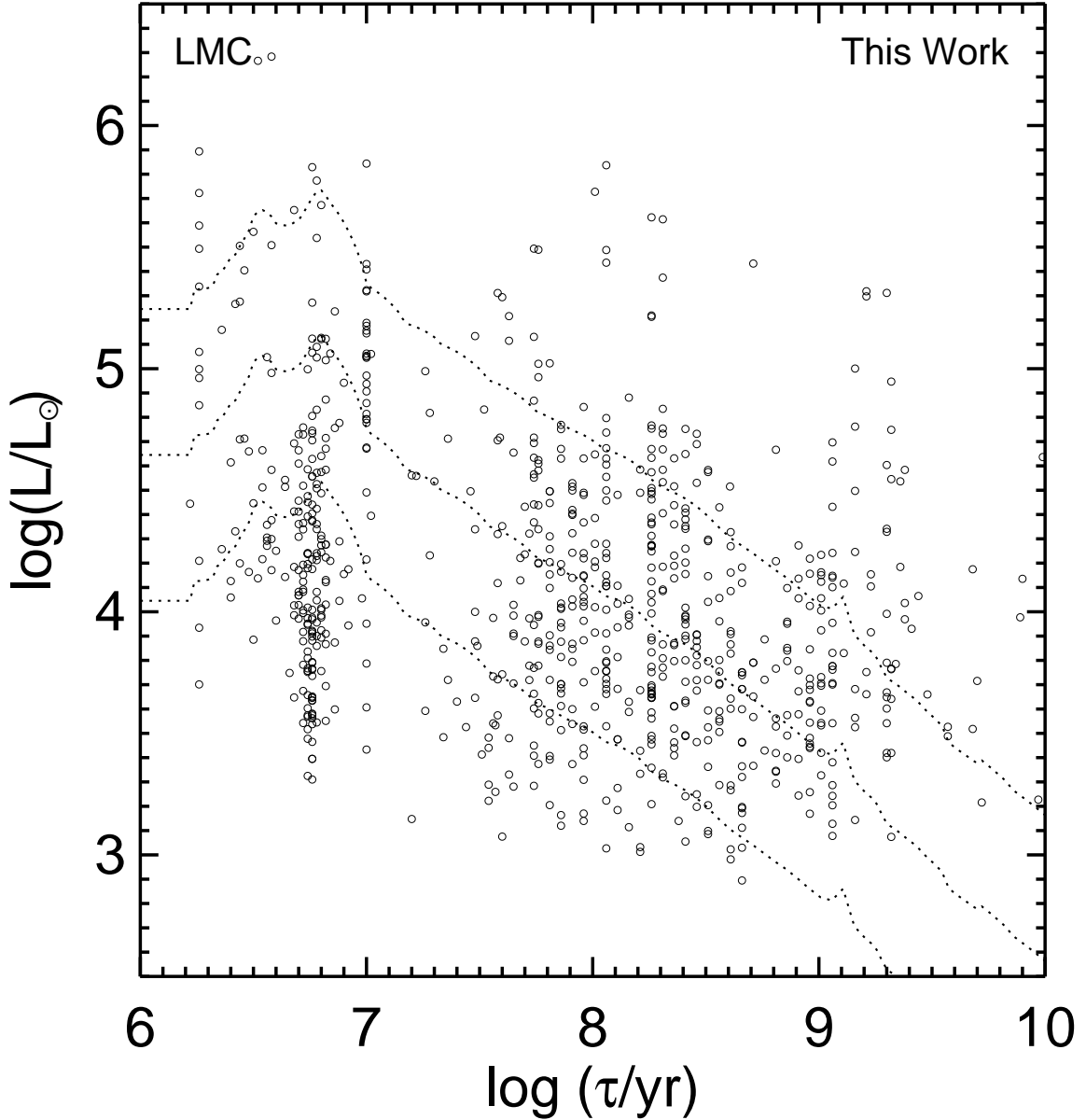}
\caption{Luminosity-age distribution for the LMC clusters from our minimum $\chi^2$ analysis. The luminosities are in the $V$ band and have been corrected for extinction. The dashed diagonal lines represent the evolutionary tracks of model clusters with initial masses of $10^4\,M_{\odot}$,
$2.5\times10^{3}\,M_{\odot}$, and $6\times10^2\,M_{\odot}$ (from top to bottom).  The vertical gap at $\tau \approx (1 \rightarrow 2)\times 10^7$~yr 
is an artifact of the age-fitting procedure, as is the pile up near $5\times10^6$~yr (see text).}
\label{fig:lt}
\end{figure}

\begin{figure}
\epsscale{0.7}
\plotone{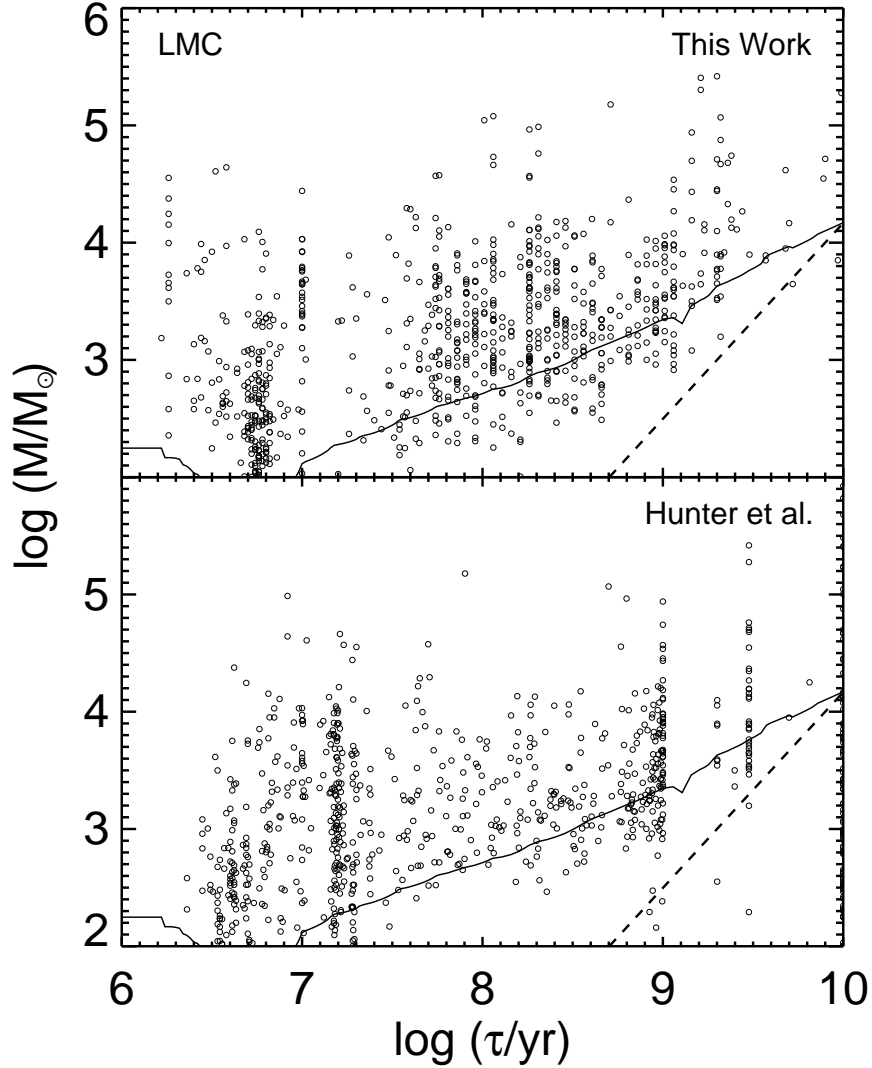}
\caption{Mass-age distribution for the LMC clusters from our minimum $\chi^2$ analysis (top panel) and the Hunter et~al.\ (2003) analysis (bottom panel). This is an equivalent representation of the data plotted in Figure~2.
The solid line in each panel shows the magnitude limit of $M_V=-4.0$, and the dashed diagonal line shows the disruption line predicted by Model~1 for $k=0.6$ and $\tau_*=8\times10^9$~yr.  See text for details.}
\label{fig:agevsmass}
\end{figure}

\begin{figure}
\epsscale{1.0}
\plotone{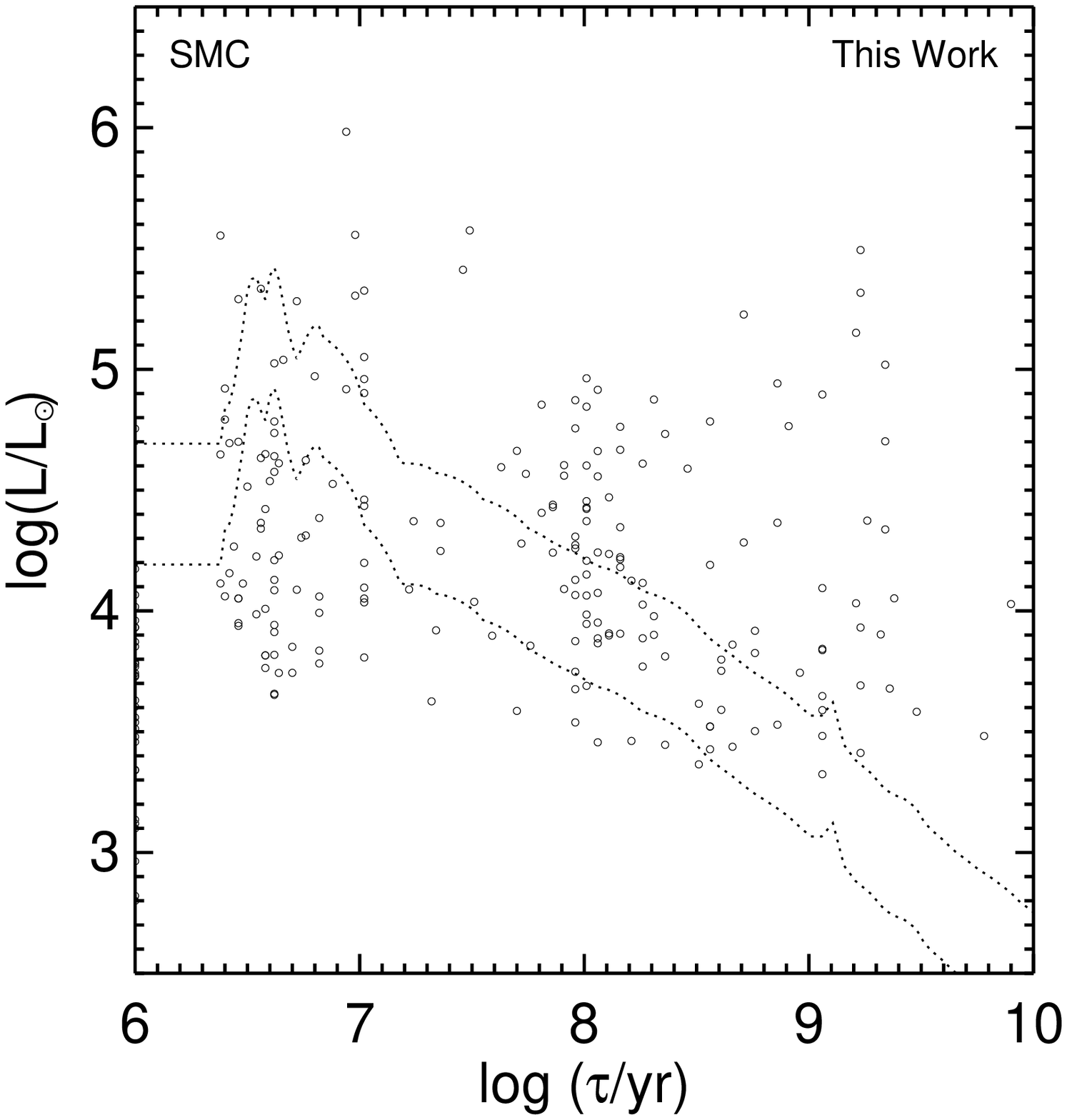}
\caption{Same as Figure~2, but for the SMC.}
\label{fig:smc_lt}
\end{figure}

\begin{figure}
\epsscale{0.9}
\plotone{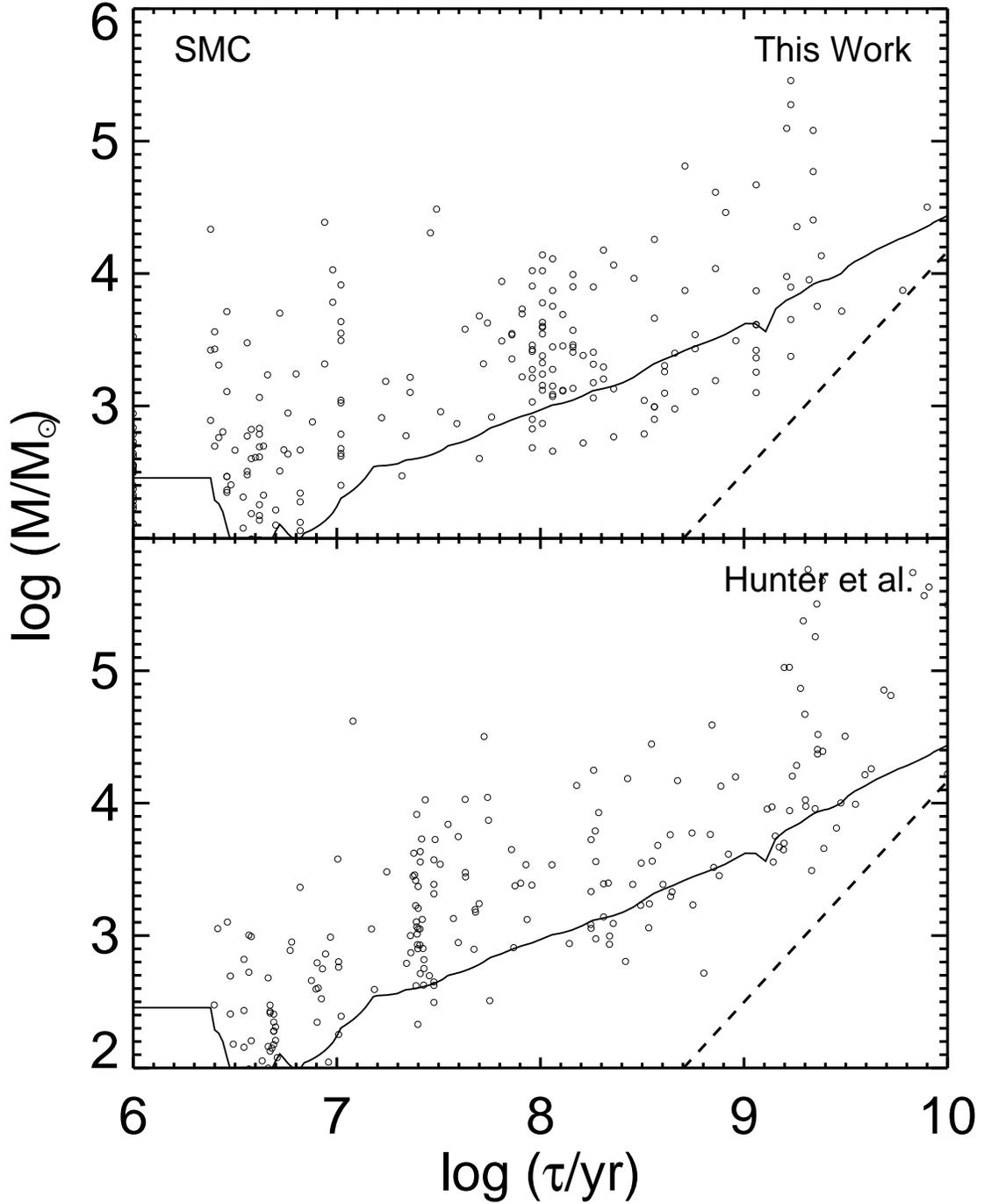}
\caption{Same as Figure~3, but for the SMC. The solid line in each panel shows the magnitude limit of $M_V=-4.5$.  The dashed line shows the disruption line predicted by Model~1 for $k=0.6$ and $\tau_*=8\times10^9$~yr.  See text for details. }
\label{fig:smc_agevsmass}
\end{figure}

\begin{figure}
\plotone{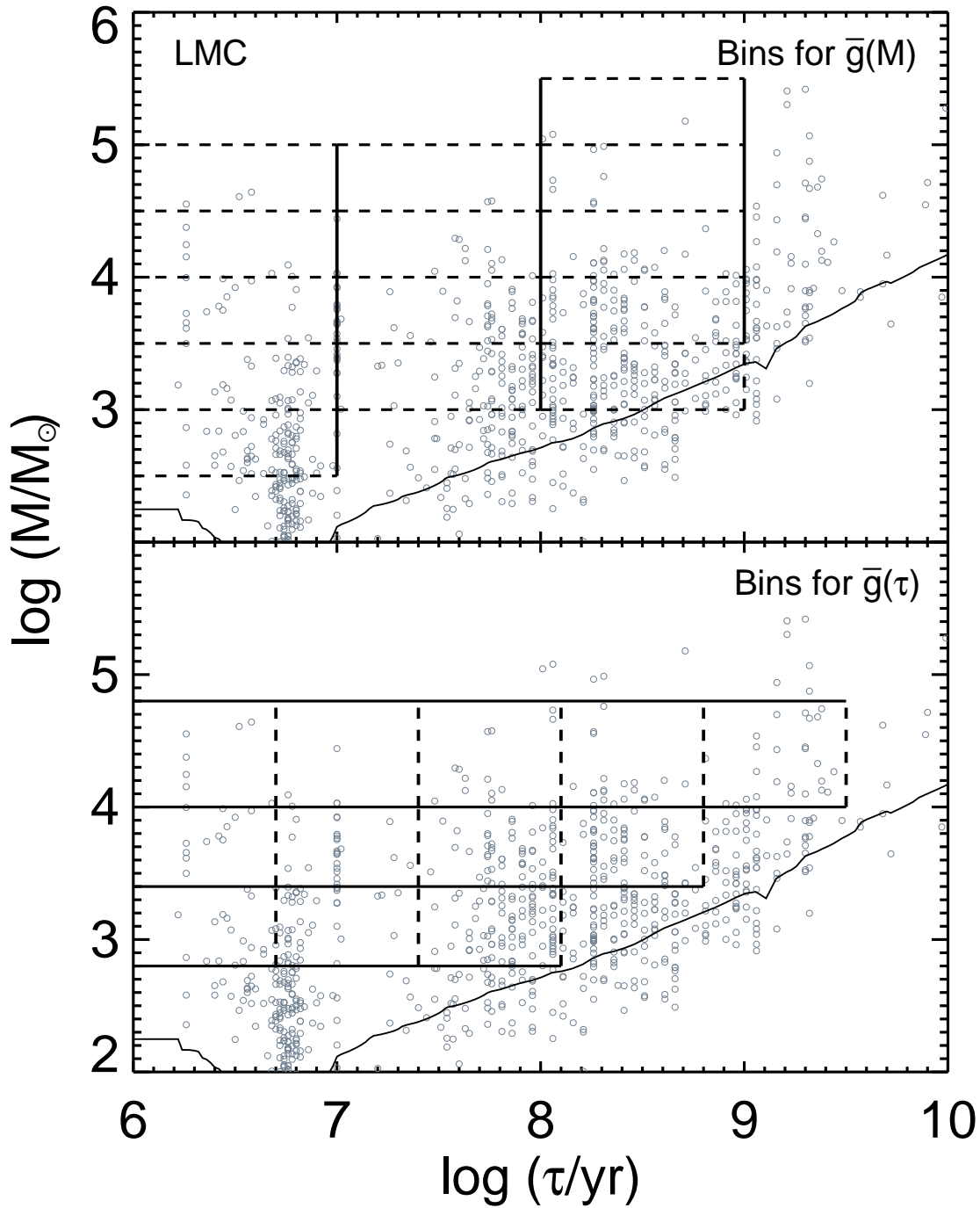}
\caption{Mass-age distribution for the LMC clusters from our minimum $\chi^2$ analysis. The dashed rectangles show the $M$--$\tau$ bins used to construct the $\bar{g}(M)$ (top panel) and $\bar{g}(\tau)$ (bottom panel) distributions presented in Section~3.2.}
\label{fig:grid}
\end{figure}

\begin{figure}
\epsscale{1.0}
\plottwo{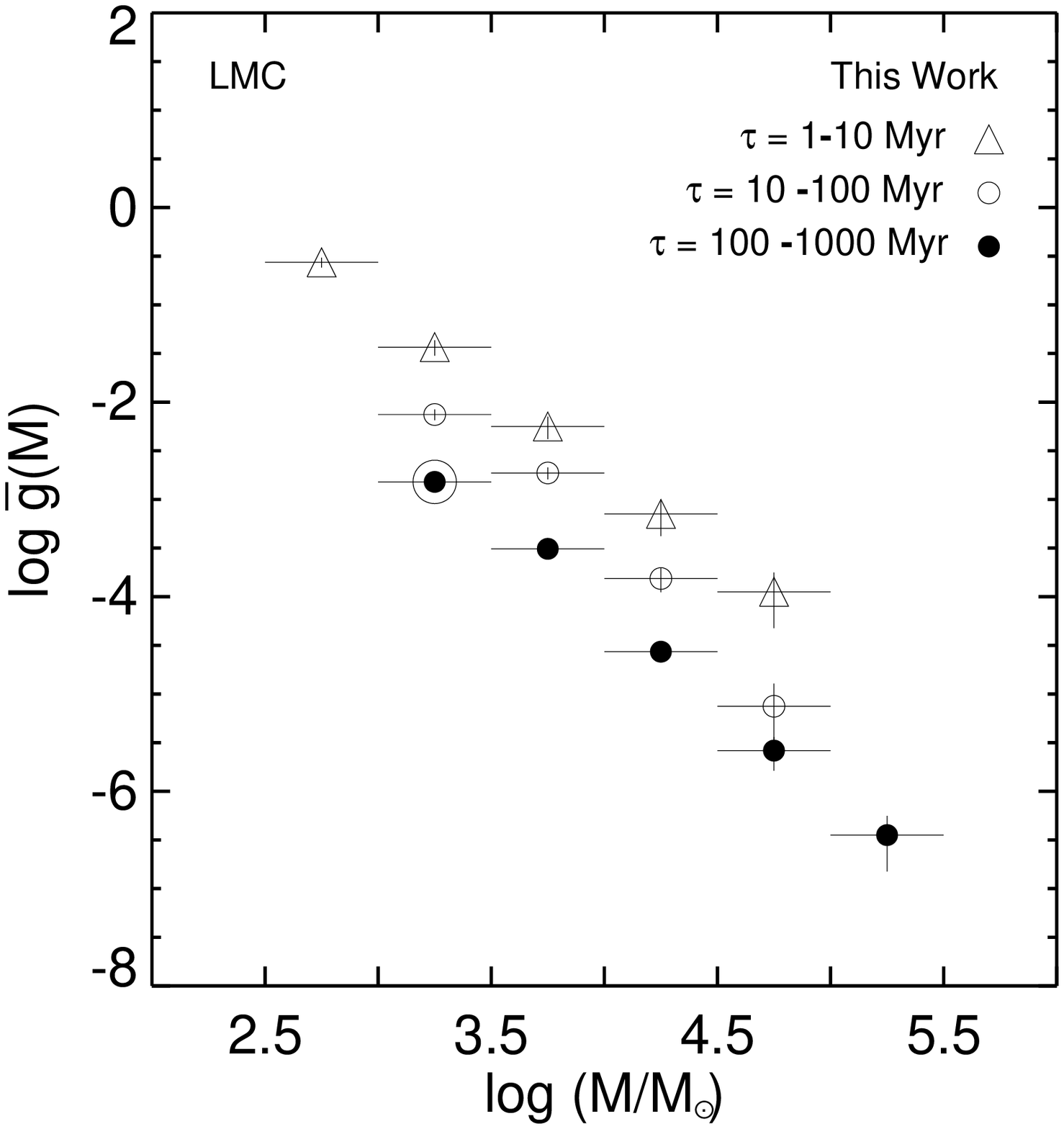}{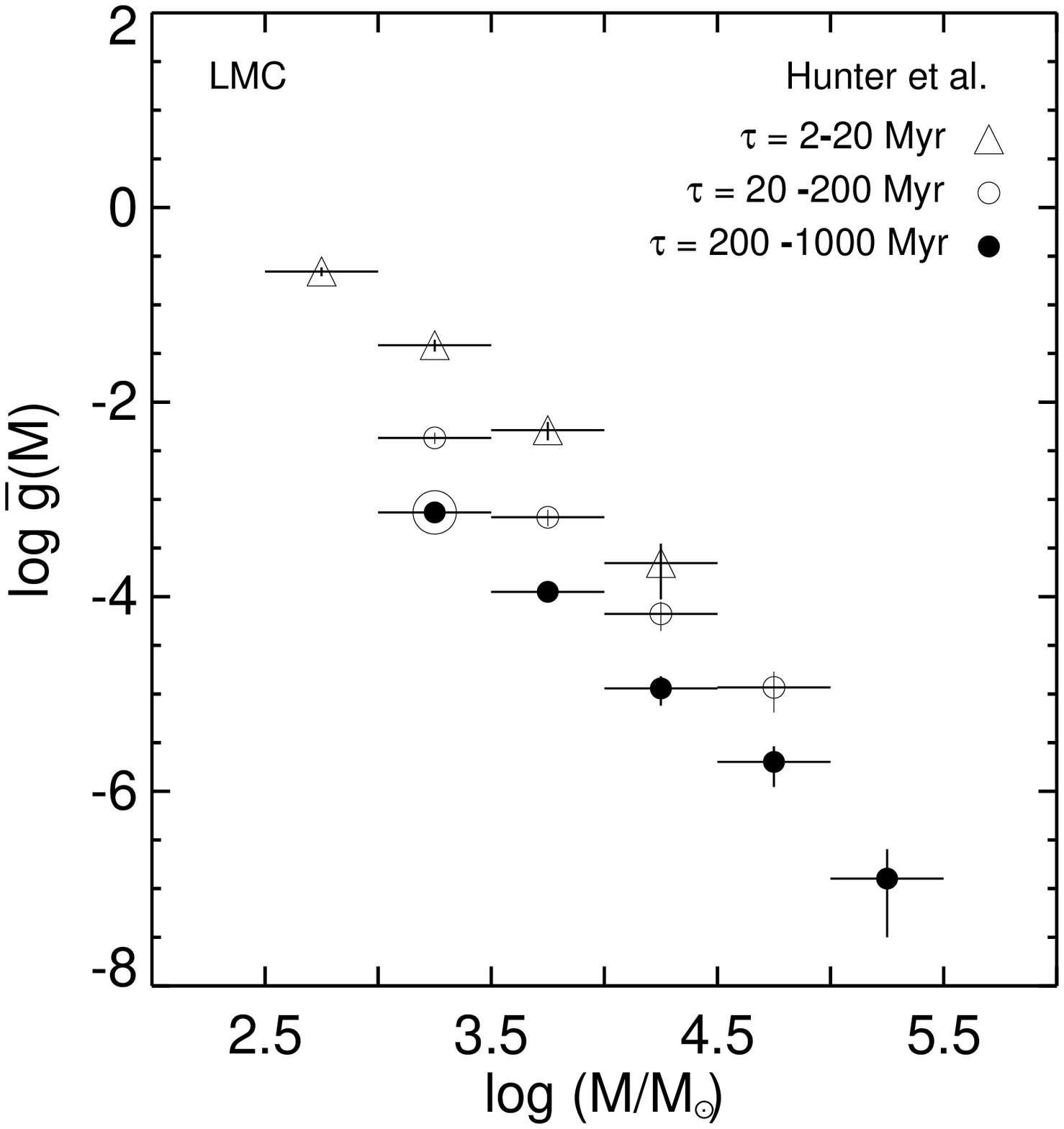}
\caption{Mass function 
of LMC clusters in the indicated intervals of age, based on our minimum $\chi^2$ analysis (left) and the Hunter et~al.\ analysis (right). The circled point contains data slightly below the $M_V=-4.0$ limit, and hence may be affected by incompleteness.
}
\label{fig:dndm}
\end{figure}

\begin{figure}
\plottwo{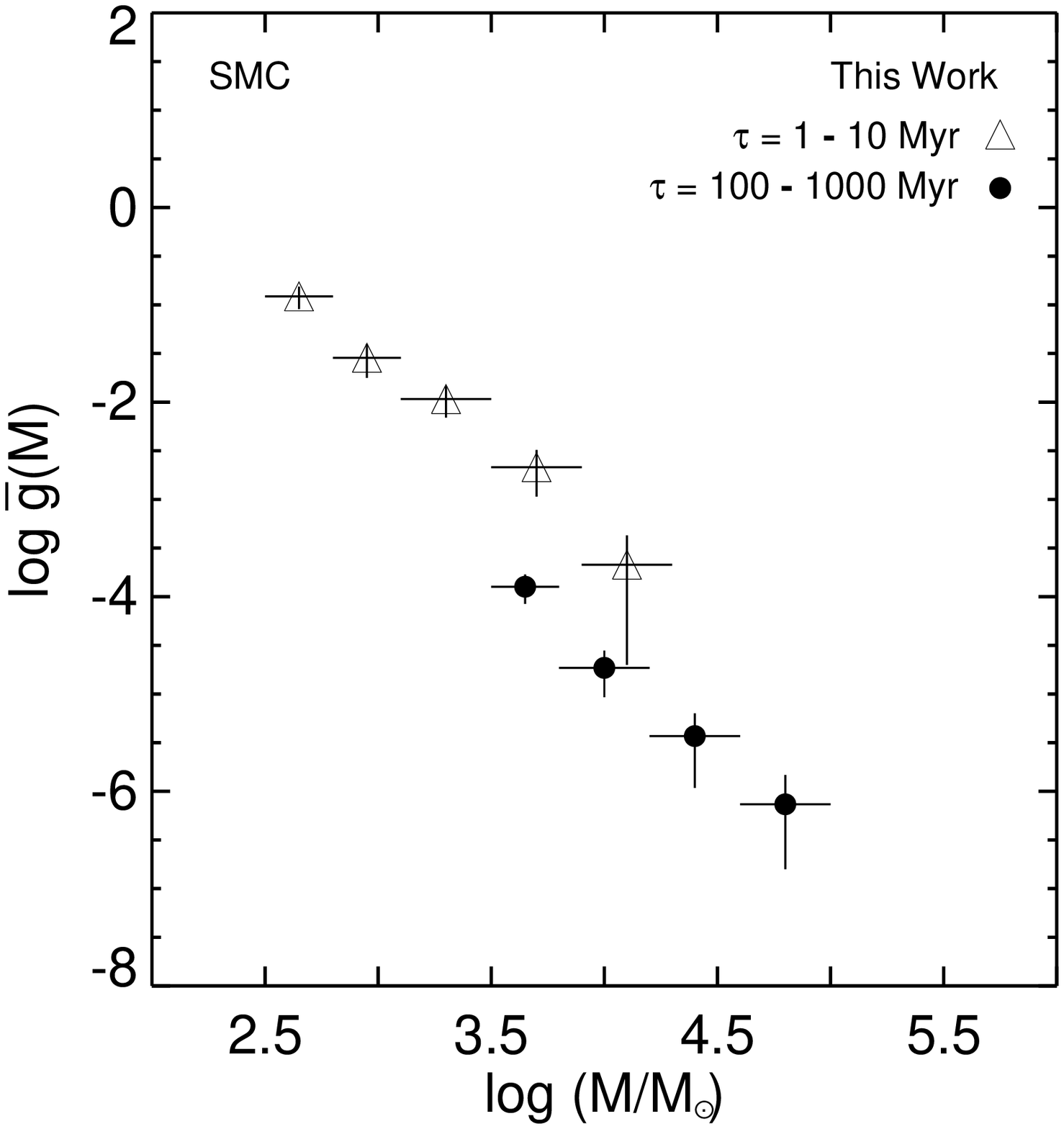}{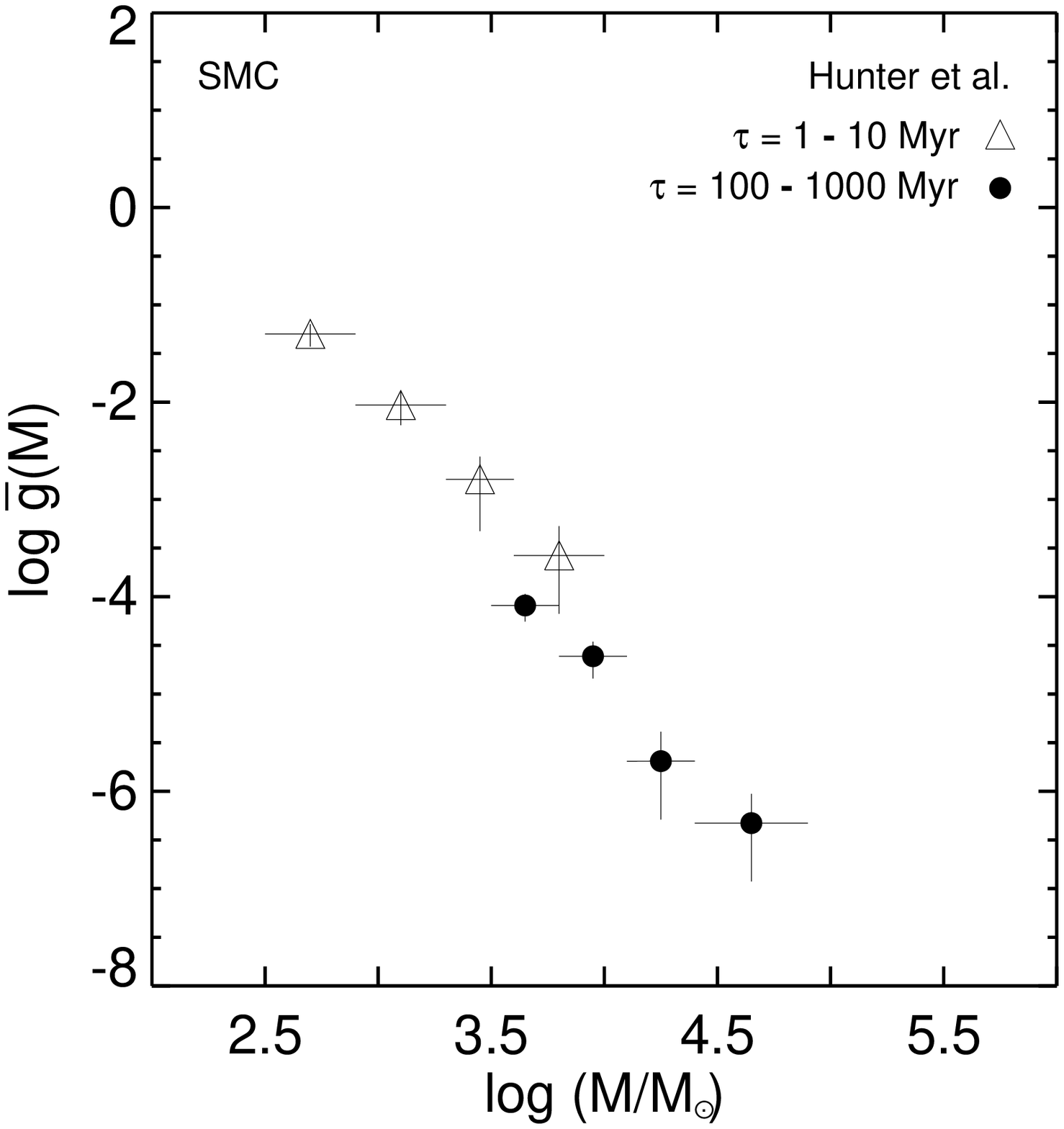}
\caption{Same as Figure~7, but for SMC clusters.}
\label{fig:smc_dndm}
\end{figure}

\begin{figure}
\plottwo{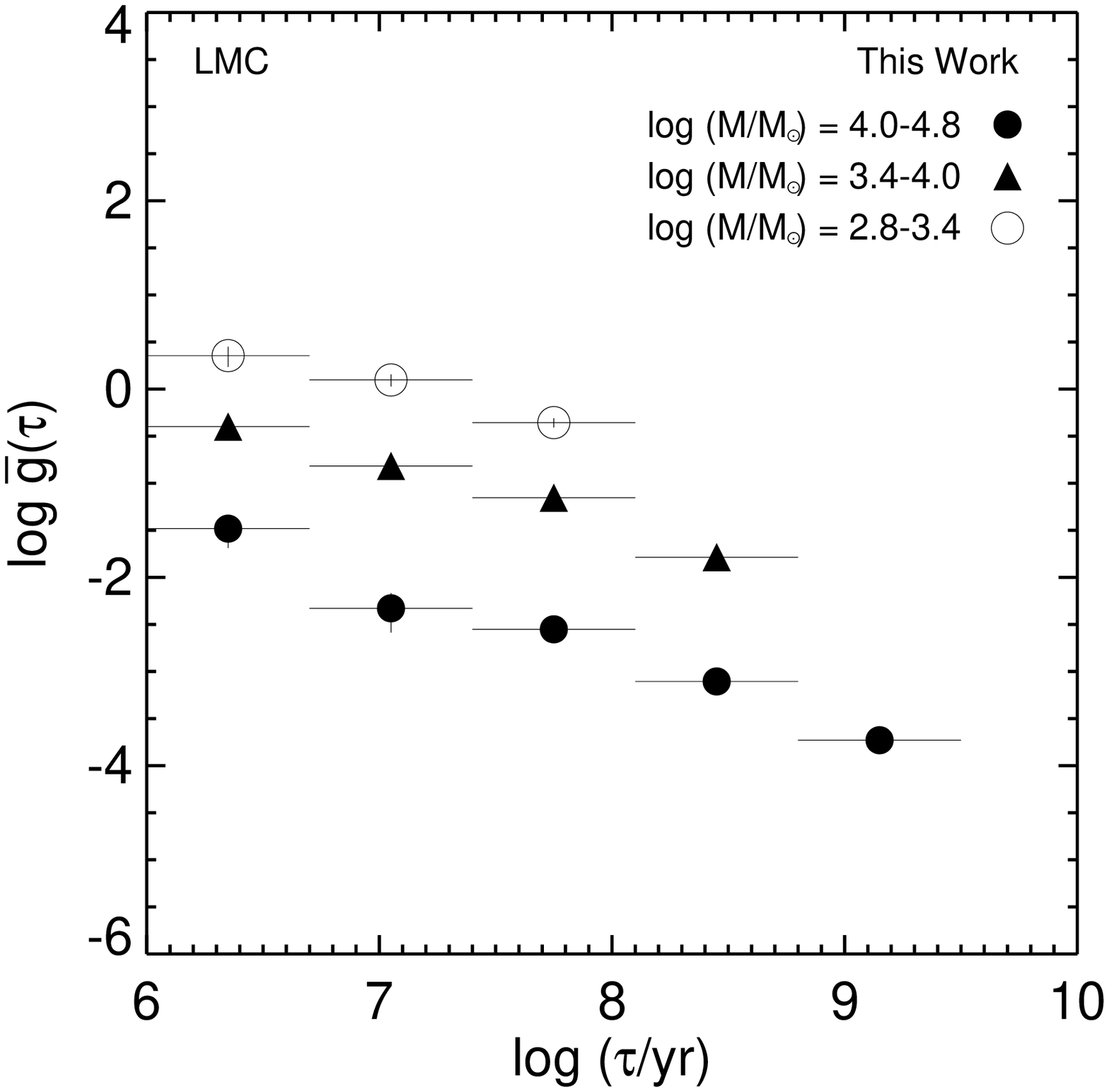}{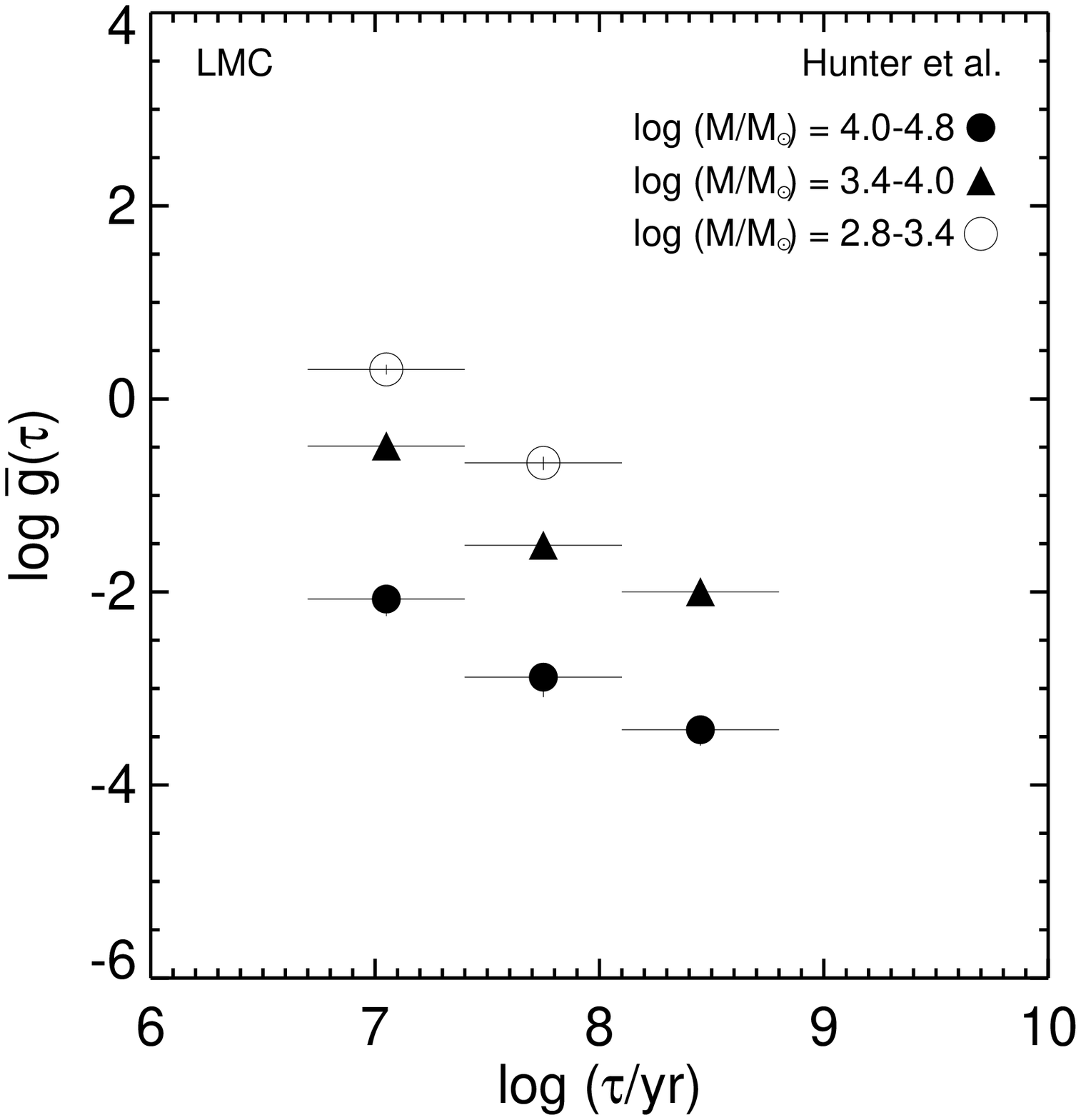}
\caption{Age distribution of LMC clusters in the indicated intervals of mass, based on our minimum $\chi^2$ analysis (left) and the Hunter et~al.\  analysis (right).}
\label{fig:smcdndt}
\end{figure}

\begin{figure}
\plottwo{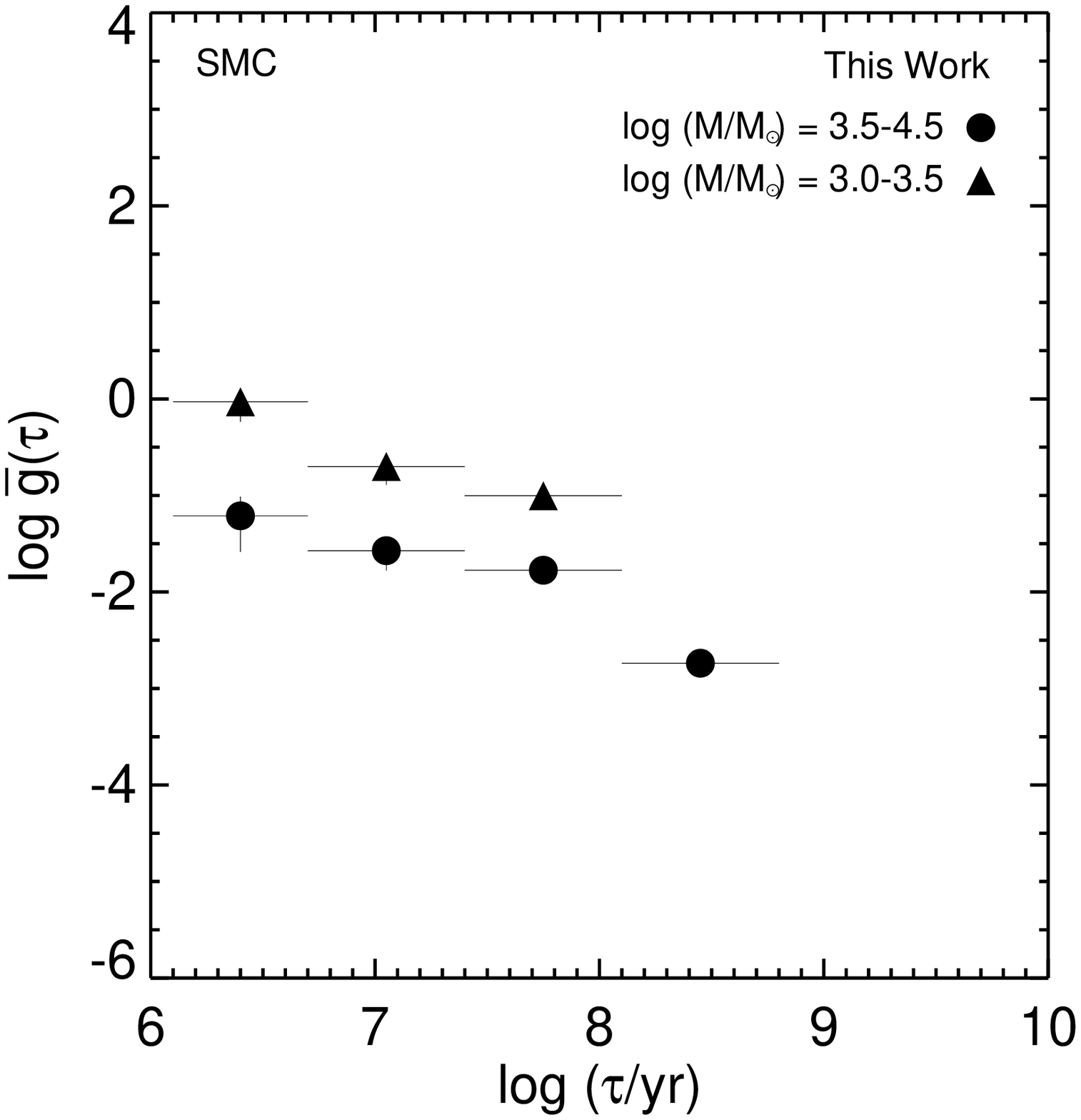}{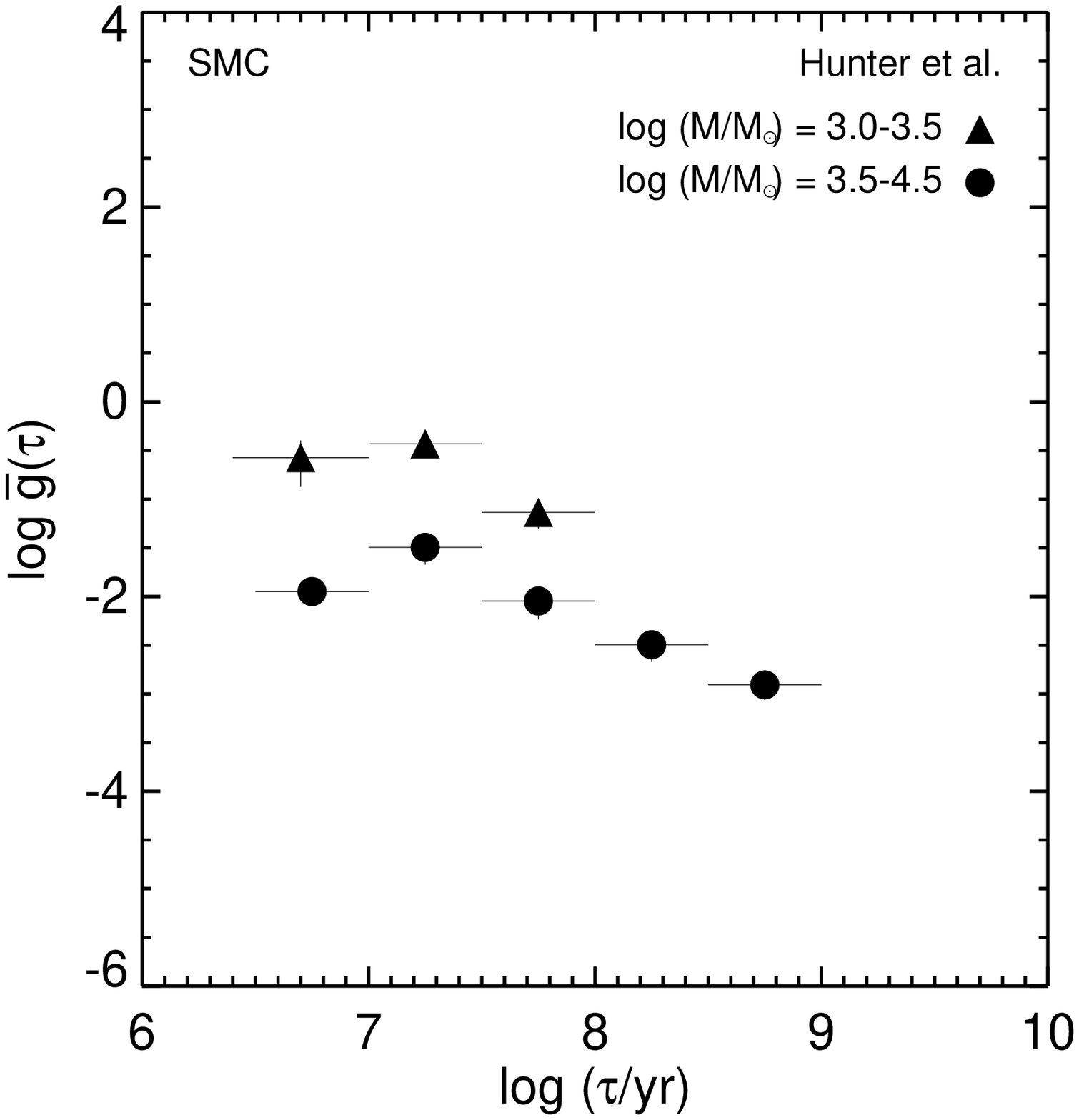}
\caption{
Same as Figure~9, but for the SMC.}
\label{fig:smcdndt}
\end{figure}

\begin{figure}
\plottwo{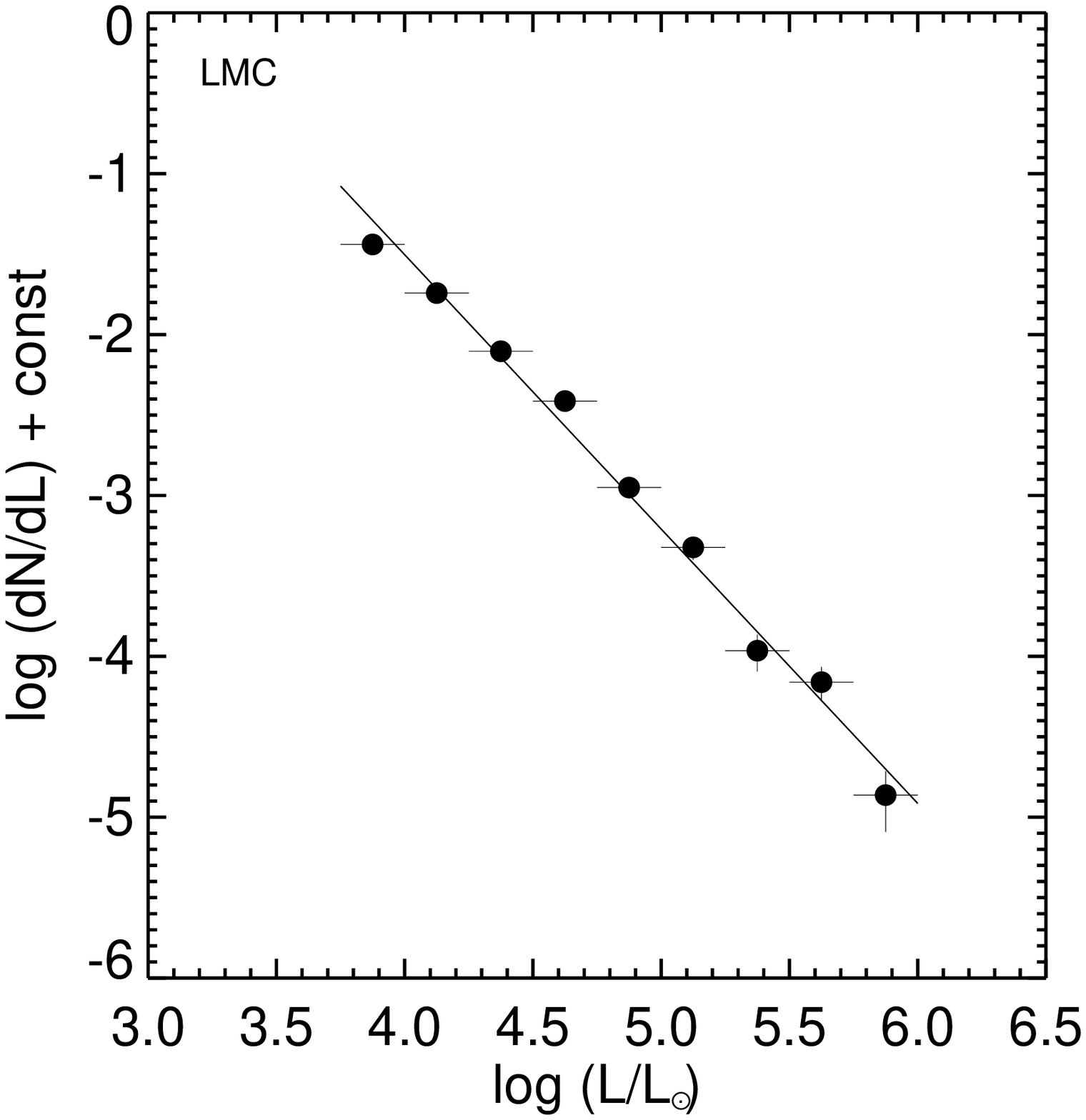}{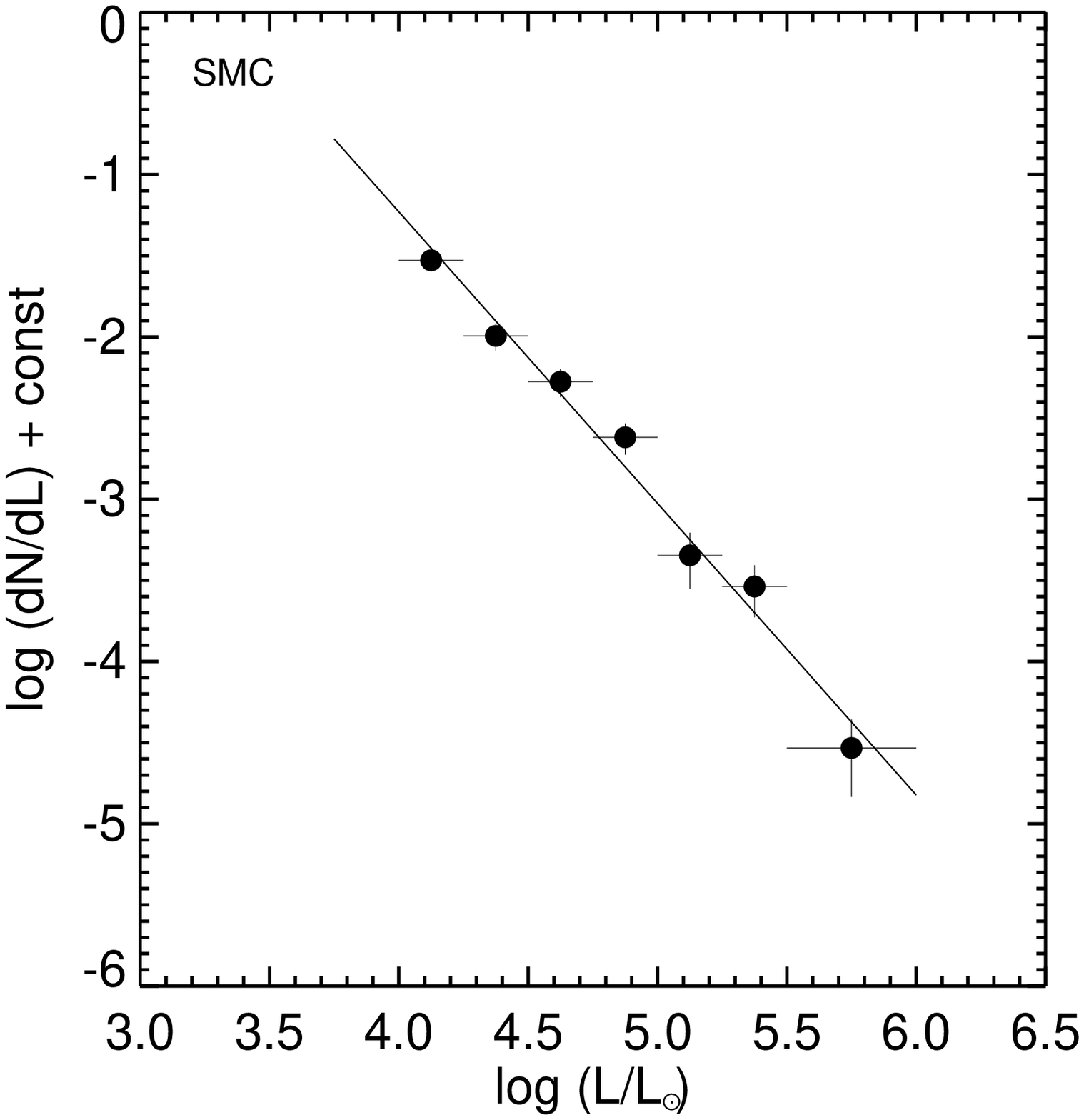}
\caption{Luminosity function of clusters in the LMC (left) and in the SMC (right) in the $V$ band. The magnitudes have been corrected for extinction.}
\label{fig:dndl}
\end{figure}

\begin{figure}
\plottwo{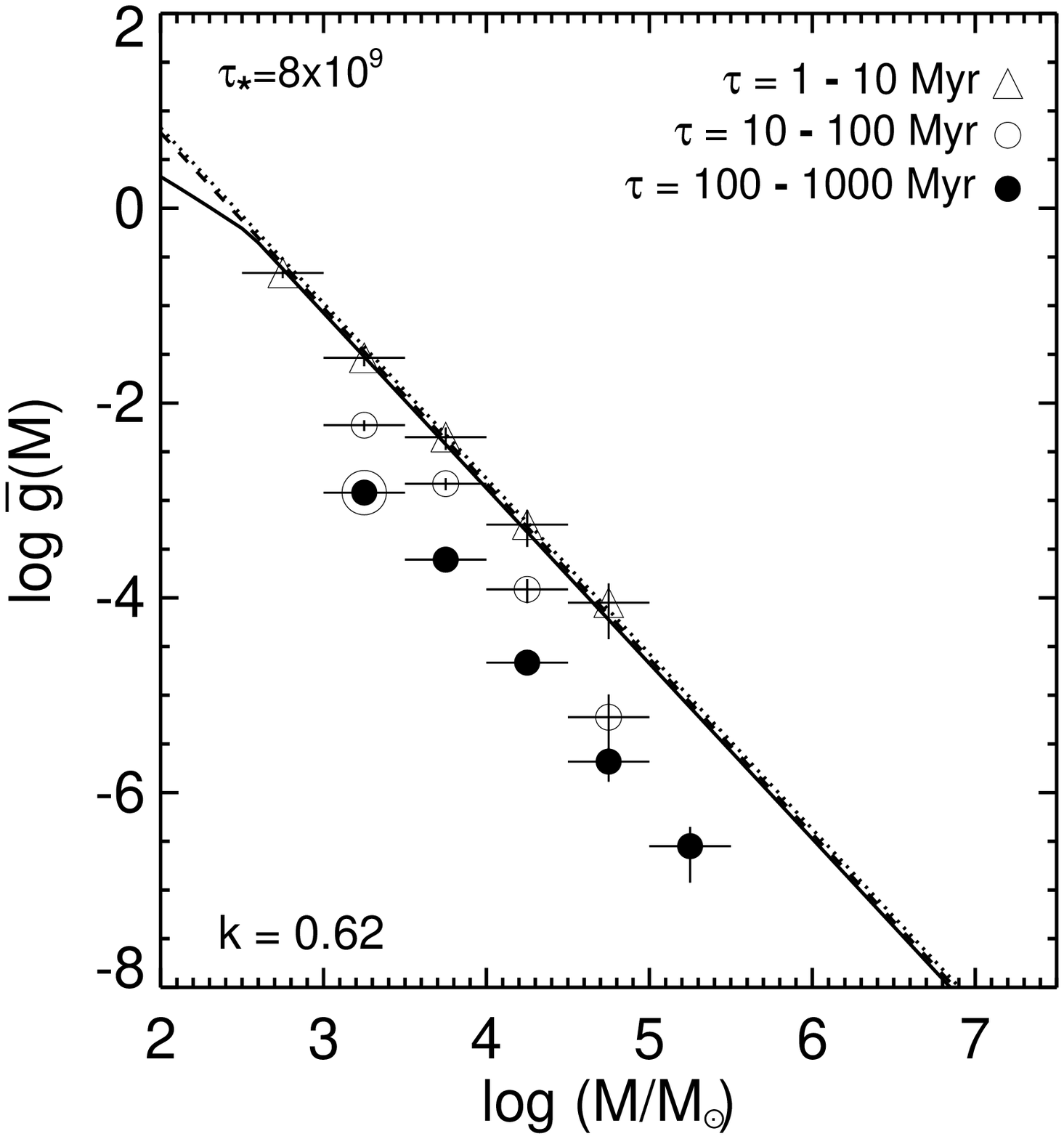}{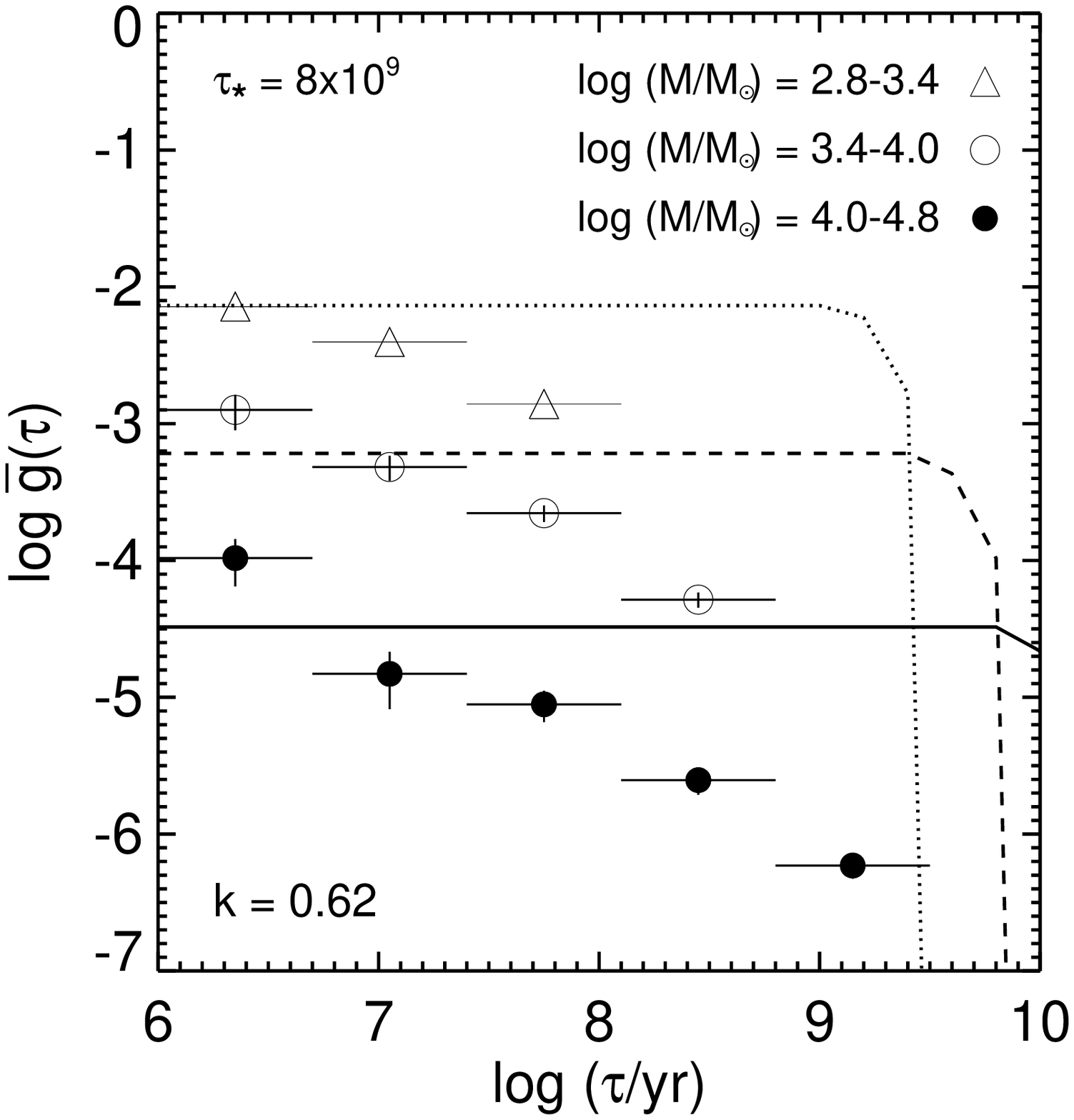}
\caption{{\bf Left:} Mass distribution averaged over the indicated intervals
of age for the LMC clusters based on our minimum $\chi^2$ analysis (data points), and for Model~1 with $k=0.62$ and $\tau_*=8\times10^9$~yr. In this model, clusters have a power-law initial mass function, a constant rate of formation, and are disrupted suddenly at an age that depends on their initial mass. The lines were computed from Equations~(B1) and (B2) in FCW09, with the solid line showing the predictions for the oldest clusters, 
and the dashed and dotted lines the predictions for the two younger age intervals. {\bf Right:} Age distribution averaged over the indicated intervals of mass for the LMC clusters based on our minimum $\chi^2$ analysis, and for Model~1 with the same parameters given above. The lines were computed from Equations (B3)--(B5) in FCW09, and normalized to the youngest data point for the lowest-mass distribution.
}
\label{fig:simagevsmass}
\end{figure}

\begin{figure}
\plotone{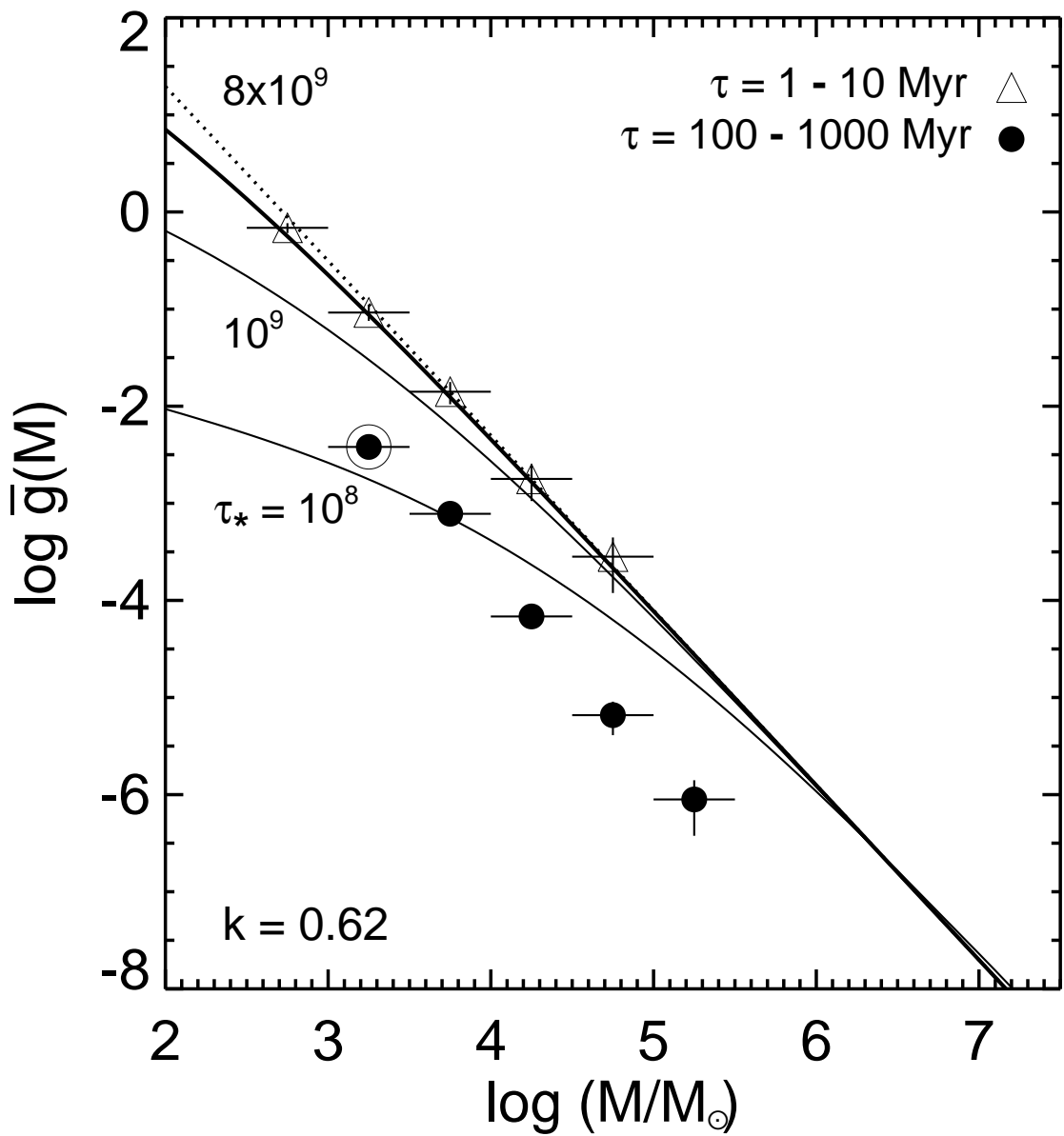}
\caption{
Mass distribution averaged over the indicated intervals of age for the LMC clusters (data points) and for Model~2 with $k=0.62$ and the labeled values of $\tau_*$. In this model, clusters have a power-law initial mass function,
a constant rate of formation, and are disrupted gradually at a rate that depend on their masses. The solid lines were computed from Equations~(B6)--(B8) in FCW09. The dotted line is $\bar{g}(M) \propto M^{-1.8}$.}
\label{fig:gmbar06}
\end{figure}

\begin{figure}
\plottwo{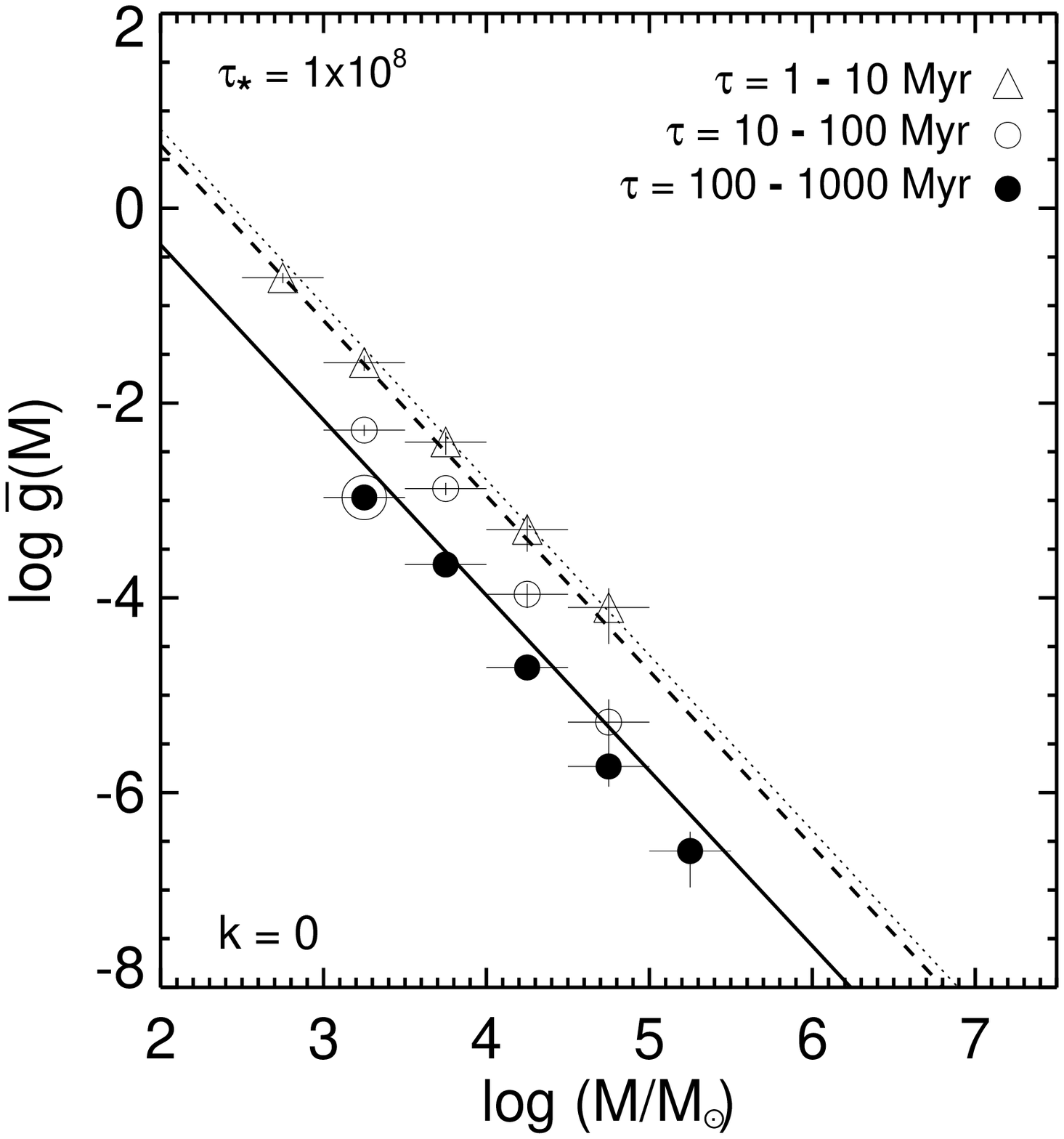}{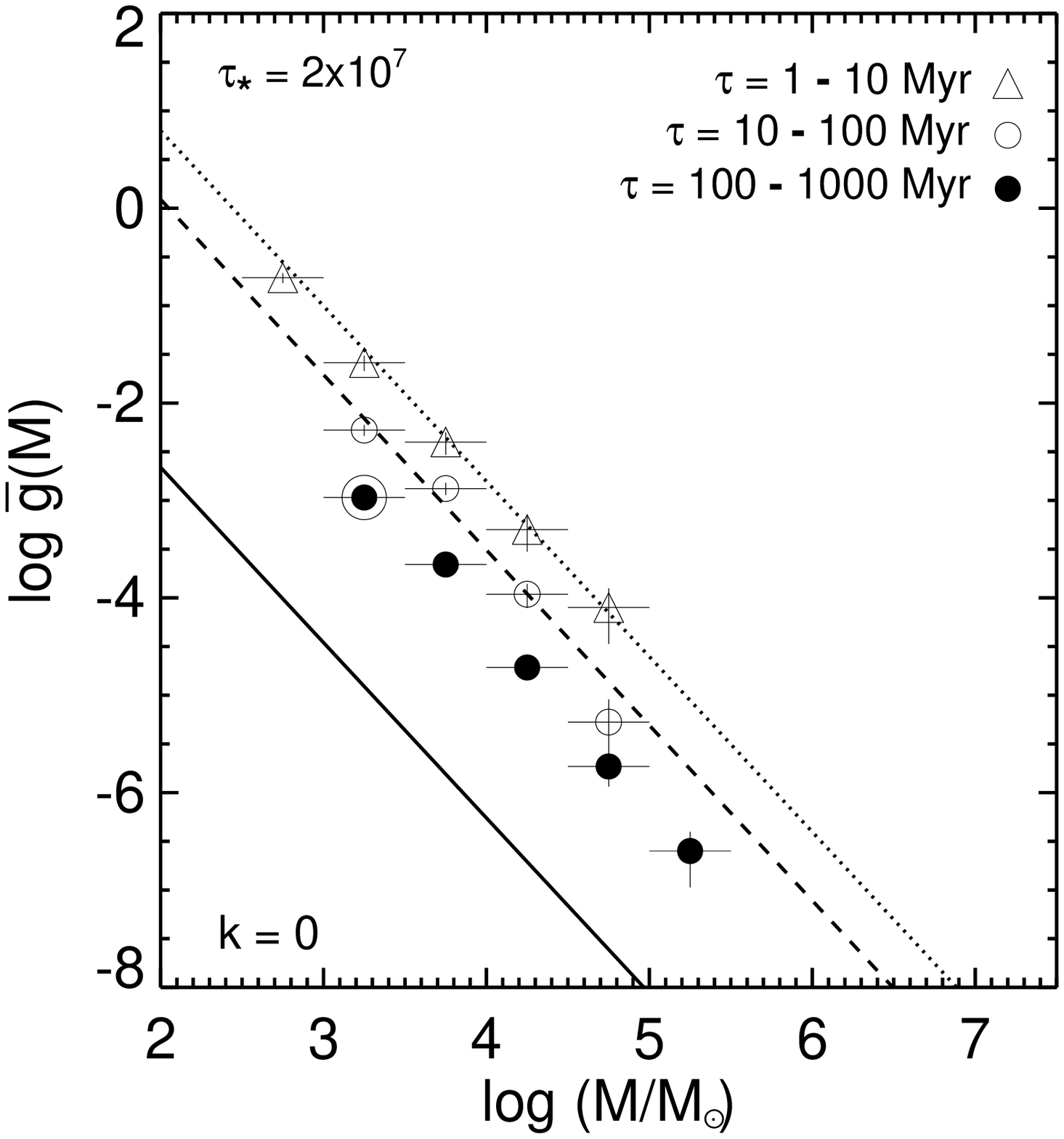}
\caption{Mass distribution averaged over the indicated intervals of age for the LMC clusters (data points) and for Model~2 with $k=0$ (lines). In this model, clusters have a power-law initial mass function, a constant rate of formation, and are disrupted gradually at a rate that does not depend on their
masses. The lines were computed for the two sets of older clusters from Equations~(B9) and (B10) in FCW09, with the indicated values of $\tau_*$.
No value of $\tau_*$ simultaneously matches the data for the two older age intervals.}
\label{fig:gmbar0}
\end{figure}

\begin{figure}
\plotone{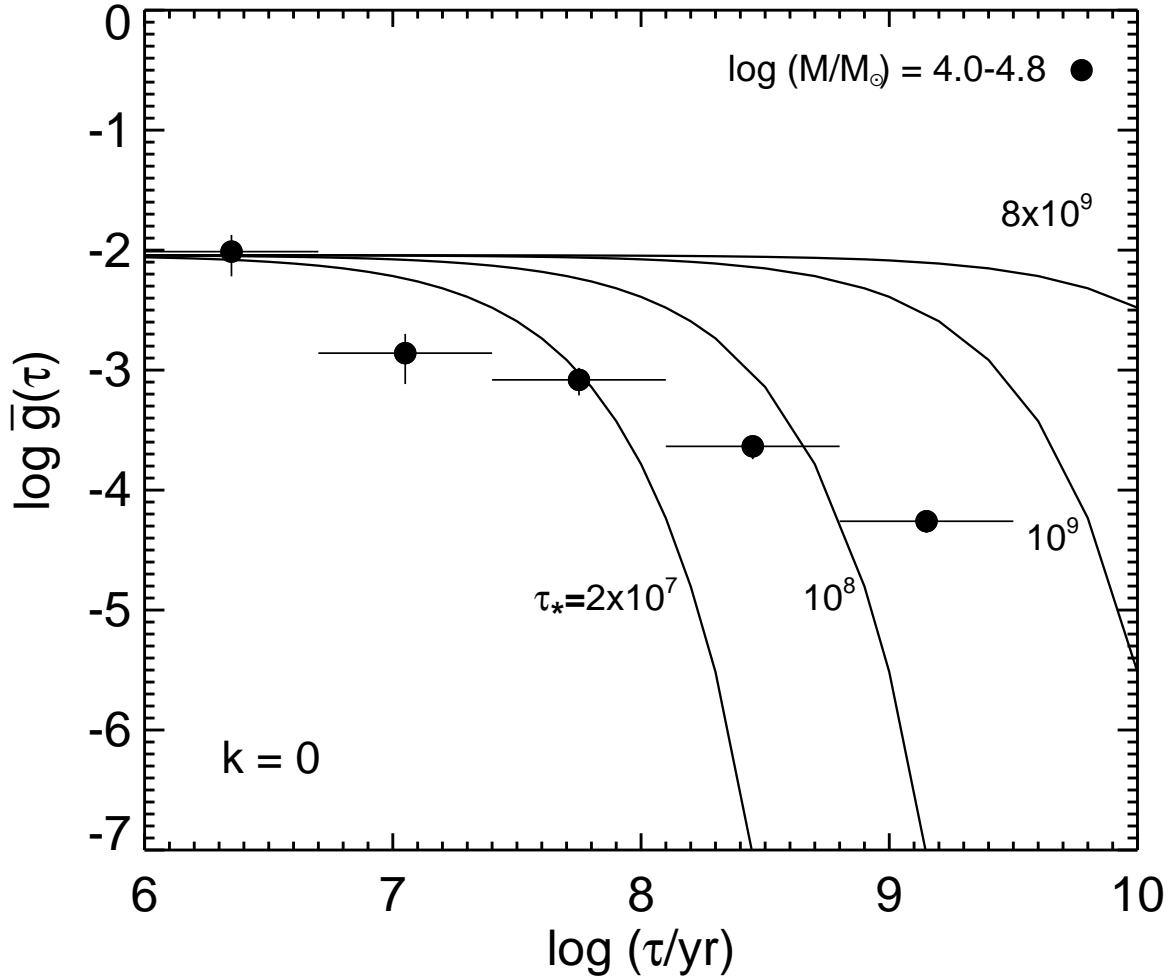}
\caption{Age distribution averaged over the indicated interval of mass for the LMC clusters (data points) and for Model~2 with $k=0$ (lines). In this model, clusters have a power-law initial mass function, a constant rate of formation, and are disrupted gradually at a rate that does not depend on their
masses. The solid lines were computed from Equations~(B5) and (B11) in FCW09, with the indicated values of $\tau_*$.}
\label{fig:gtbar0}
\end{figure}

\begin{figure}
\plottwo{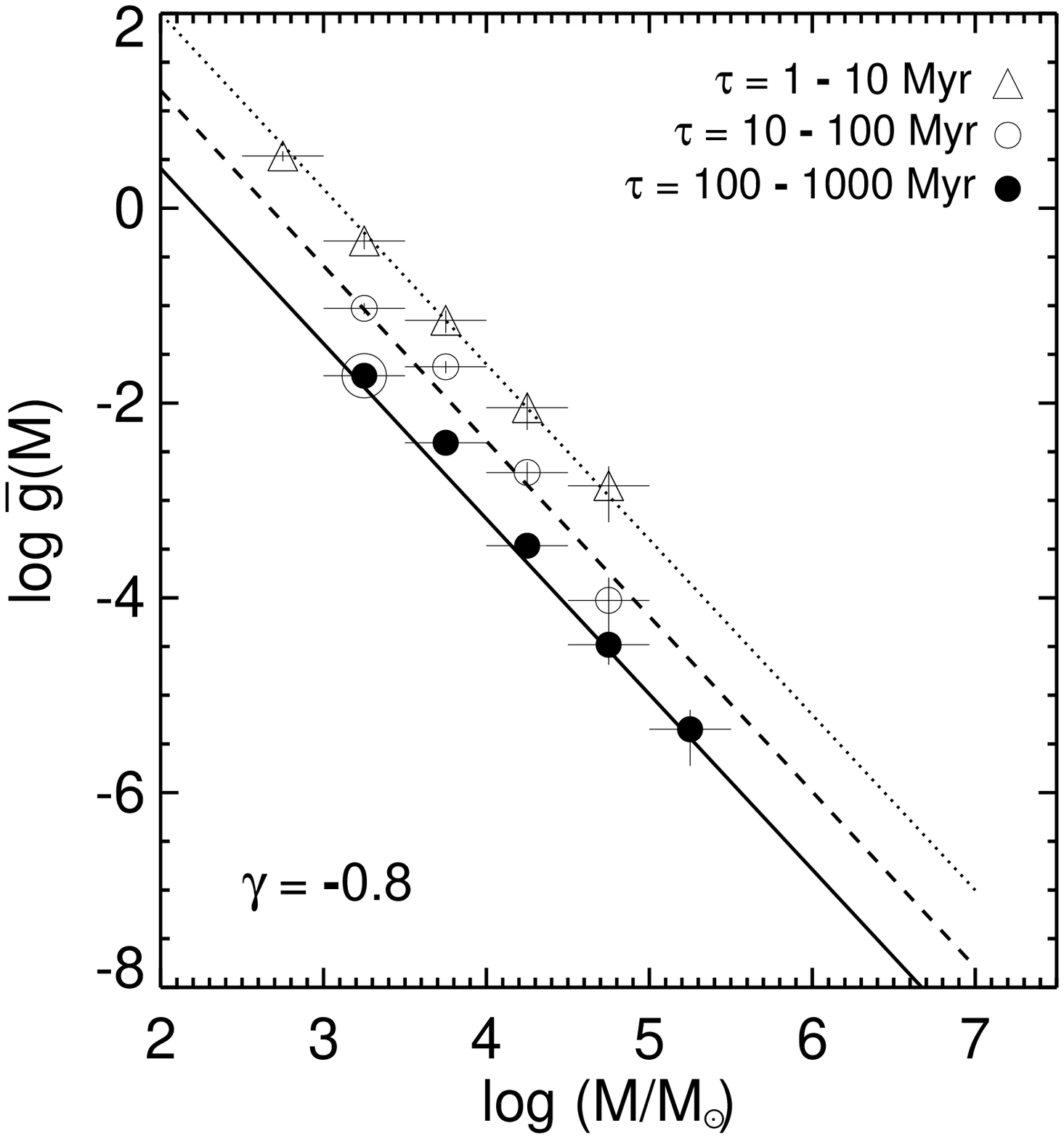}{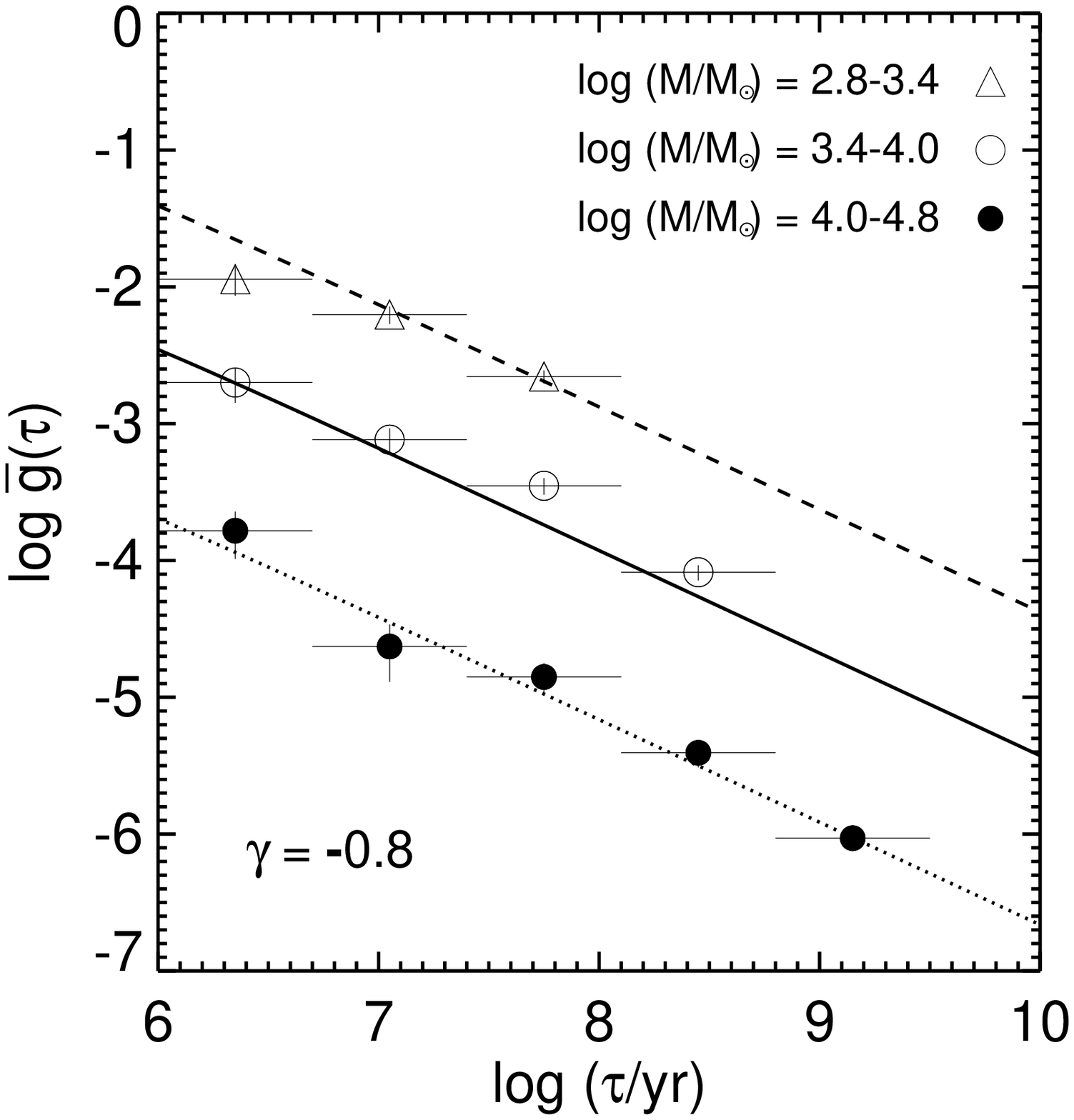}
\caption{{\bf Left:} Mass distribution averaged over the indicated intervals of age for the LMC clusters based on our minimum $\chi^2$ analysis (data points), and for Model~3 with $\gamma=-0.8$ (lines), using Equations (B12) and (B13) from FCW09. In this model, clusters have a power-law initial mass function, a constant rate of formation, and are disrupted gradually, at a rate
that is {\em in}dependent of their masses. {\bf Right:} The age distribution averaged over the indicated intervals of mass for the LMC clusters plotted for
Model~3 with the same parameters as the mass distribution, using Equations (B5) and (B16) in FCW09.}
\label{fig:gmnew}
\end{figure}

\begin{figure}
\plottwo{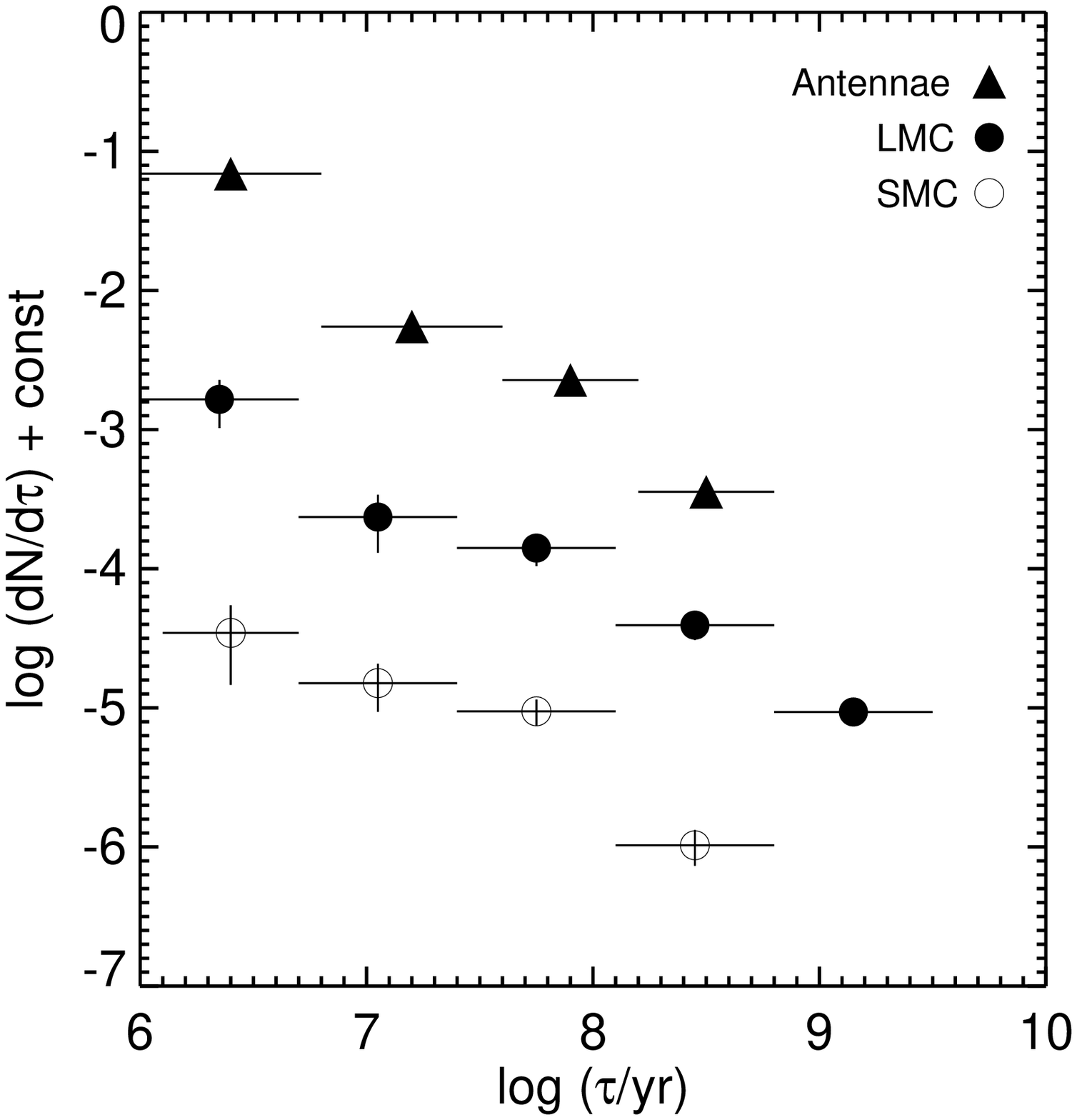}{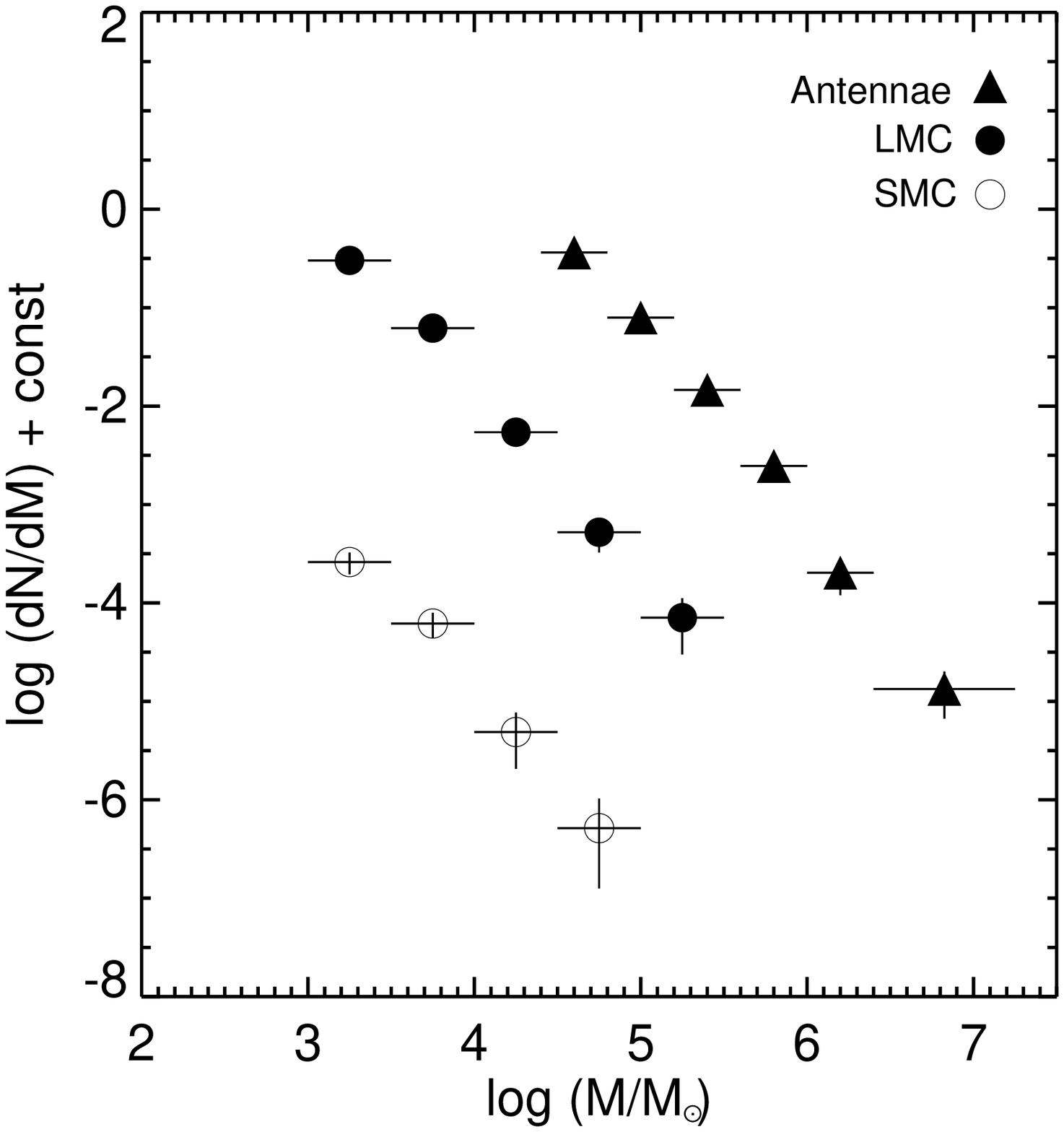}
\caption{Age and mass distributions in the Antennae, LMC, and SMC.  }
\label{fig:dndm}
\end{figure}

\end{document}